\input epsf 
\documentstyle[preprint,eqsecnum,aps]{revtex}

\begin{document}

\count255=\time\divide\count255 by 60 \xdef\hourmin{\number\count255}
  \multiply\count255 by-60\advance\count255 by\time
 \xdef\hourmin{\hourmin:\ifnum\count255<10 0\fi\the\count255}

%\draft
\preprint{\vbox{\hbox{WM-00-101}\hbox{JLAB-THY-00-04}}}

\title{Maximal Neutrino Mixing from a Minimal Flavor Symmetry}

\author{Alfredo Aranda$^a$\footnote{fefo@physics.wm.edu}, 
Christopher D.  Carone$^a$\footnote{carone@physics.wm.edu}, 
and Richard F. Lebed$^b$\footnote{lebed@jlab.org}}

\vskip 0.1in

\address{$^a$Nuclear and Particle Theory Group, Department of
Physics, College of William and Mary, Williamsburg, VA 23187-8795\\
$^b$Jefferson Lab, 12000 Jefferson Avenue, Newport News, VA 23606}
\vskip .1in
\date{January, 2000}
\vskip .1in

\maketitle
\tightenlines
\thispagestyle{empty}

\begin{abstract}
We study a number of models, based on a non-Abelian discrete group,
that successfully reproduce the simple and predictive Yukawa
textures usually associated with U(2) theories of flavor.  These models
allow for solutions to the solar and atmospheric neutrino problems 
that do not require altering successful predictions for the charged fermions
or introducing sterile neutrinos.  Although Yukawa matrices 
are hierarchical in the models we consider, the mixing between second- and
third-generation neutrinos is naturally large.  We first present a
quantitative analysis of a minimal model proposed in earlier work, 
consisting of a global fit to fermion masses and mixing angles, including 
the most important renormalization group effects.  We then propose two new
variant models:  The first reproduces all important features of the 
SU(5)$\times$U(2) 
unified theory with neither SU(5) nor U(2).  The second demonstrates that
discrete subgroups of SU(2) can be used in constructing viable 
supersymmetric theories of flavor without scalar universality
even though SU(2) by itself cannot.
\end{abstract}


\newpage
\setcounter{page}{1}

\section{Introduction}\label{sec:intro}

It is possible that the observed hierarchy of fermion masses and 
mixing angles originates from the spontaneous breakdown of a new 
symmetry $G_f$ that acts horizontally across the three standard model 
generations.  Ideally, all Yukawa couplings except that of 
the top quark are forbidden by $G_f$ invariance at high energies; the 
remaining ones are generated when a set of fields $\phi$ that transform 
nontrivially under $G_f$ develop vacuum expectation values (vevs).  A hierarchy 
in couplings is obtained if $G_f$ is broken sequentially at energy 
scales $\mu_i$ through a series of nested subgroups $H_i$ , such that    
\begin{equation}
G_f \stackrel{\mu_1}{\longrightarrow} H_1 
\stackrel{\mu_2}{\longrightarrow} H_2 
\stackrel{\mu_3}{\longrightarrow} \cdots \,\,\,\mbox{for}\,\,\,
\mu_1 > \mu_2 > \mu_3 \cdots  \,\,\, .
\end{equation}
At each stage of the symmetry breaking there is an associated small
dimensionless parameter $\langle \phi_i \rangle/M_f$, where $\phi_i$ 
is a `flavon' field whose vev is responsible for the 
breaking $H_{i-1}\rightarrow H_{i}$, and where $M_f$ is the 
ultraviolet cutoff of the $G_f$-invariant effective theory.  
The ratios $\phi_i/M_f$ appear in higher-dimension operators
that contribute to Yukawa couplings in the low-energy theory.
For example, the superpotential term
\begin{equation}
\frac{1}{M_f} Q_3 H_D \phi_b D_3
\end{equation}
leads to a bottom quark Yukawa coupling of order $\langle \phi_b \rangle/M_f$.  
The most general set of operators involving the fields of the minimal 
supersymmetric standard model (MSSM) and the $\phi$ fields must provide for 
Yukawa textures that are phenomenologically viable.  If flavor universality of 
scalar superpartner masses is not simply a consequence of the 
mechanism by which supersymmetry 
breaking is mediated~\cite{DN,randall,luty,dim}, 
then a successful model must also explain why 
these scalars do not contribute to flavor-changing
neutral current processes at unacceptable levels.

Models with horizontal symmetries have been proposed with $G_f$
either gauged or global, continuous or discrete, Abelian or non-Abelian, 
or some appropriate combination thereof~\cite{abelian,nonabelian}.  
Abelian flavor symmetries have
been used successfully to explain the absence of supersymmetric 
flavor-changing processes by aligning the fermion and sfermion 
mass matrices~\cite{abelian}.  
However, the freedom to choose a number of new U(1) charges 
for each MSSM matter field represents so much freedom that these models 
seem {\em ad hoc}, at least from a low-energy point of view.  
Non-Abelian symmetries are more restrictive, as the Yukawa matrices 
generally decompose into a smaller number of irreducible $G_f$ 
representations. Thus, it is not unreasonable to expect that minimal 
models exist that are both successful and aesthetically compelling. This 
is the primary motivation for the current work.

In non-Abelian flavor models, the existence of three generations of 
matter fields, the heaviness of the top quark, and the absence of 
supersymmetric flavor-changing processes together suggest 
a {\bf 2}$\oplus${\bf 1} representation structure for the MSSM matter
fields.  With this choice it is not only possible to distinguish the 
third generation, but also to achieve an exact degeneracy between 
superparticles of the first two generations when $G_f$ is unbroken.  In 
the low-energy theory, this degeneracy is lifted by the same small 
symmetry-breaking parameters that determine the light fermion Yukawa 
couplings, so that FCNC effects remain adequately suppressed, even with 
superparticle masses less than a TeV.  

A particularly elegant model of this type considered in the literature 
assumes the continuous, global symmetry $G_f=$U(2)~\cite{u21,u22,u23}.
Quarks and leptons are assigned to {\bf 2}$\oplus${\bf 1} representations, 
so that in tensor notation one may represent the three generations of any 
matter field by $F^a+F^3$, where $a$ is a U(2) index, and $F$ is $Q$, 
$U$, $D$, $L$, or $E$.  A set of flavons is introduced consisting of 
$\phi_a$, $S_{ab}$, and $A_{ab}$, where $\phi$ is a U(2) doublet, and  
$S$($A$) is a symmetric (antisymmetric) U(2) triplet (singlet).  The 
doublet and triplet flavons acquire the vevs
\begin{equation}
\frac{\langle\phi\rangle}{M_f} = \left(\begin{array}{c} 0 
\\ \epsilon\end{array}\right),
\,\,\,\,\,\mbox{ and }\,\,\,\,\,
\frac{\langle S\rangle}{M_f} = \left(\begin{array}{cc} 0 & 0 
\\ 0 & \epsilon \end{array} 
\right),
\end{equation}
the most general set of nonvanishing entries consistent with an unbroken 
U(1) symmetry that rotates all first generation-fields by a phase.  
This residual U(1) symmetry is broken at a somewhat lower scale by the flavon $A$
\begin{equation}
\frac{\langle A\rangle}{M_f}=
\left(\begin{array}{cc} 0 & \epsilon' \\ -\epsilon' & 0 \end{array}
\right) \,\,\, ,
\end{equation}
where $\epsilon'<\epsilon$.  Thus, the sequential breaking
\begin{equation}
U(2)\stackrel{\epsilon}{\longrightarrow} U(1) 
\stackrel{\epsilon'}{\longrightarrow} nothing,
\end{equation}
yields a Yukawa texture for the down quarks, for example, of the form
\begin{equation}\label{eq:ydown}
Y_D \approx \left(\begin{array}{ccc}  0 & d_1\epsilon' & 0 \\
-d_1\epsilon' & d_2\epsilon & d_3\epsilon \\
0 & d_4 \epsilon & d_5 \end{array}\right) \xi \,\,\, ,
\label{eq:yd}
\end{equation}
where $d_1,\ldots ,d_5$ are $O(1)$ coefficients.   With the
choice $\epsilon\approx 0.02$ and $\epsilon'\approx 0.004$,
this texture achieves the correct hierarchy in down quark mass 
eigenvalues and gives contributions of the appropriate size
to entries of the Cabibbo-Kobayashi-Maskawa (CKM) matrix. 
The $O(1)$ coefficients may be determined from a global fit, 
as in Ref.~\cite{u23}.   The ratio $m_b/m_t$ is assumed to be 
unrelated to U(2) symmetry breaking, and is simply put into the 
low-energy theory by hand.  This is accomplished by choosing
the free parameter $\xi$ in Eq.~(\ref{eq:ydown}).

While the form of $Y_D$ is viable, U(2) symmetry by itself 
cannot explain the differences between the hierarchies within $Y_D$ and 
$Y_U$.  Quark mass ratios renormalized at the grand unified scale 
are given approximately by~\cite{arason}
\begin{equation}
m_d :: m_s :: m_b = \lambda^4 :: \lambda^2 :: 1, 
\end{equation} 
while
\begin{equation}
m_u :: m_c :: m_t = \lambda^8 :: \lambda^4 :: 1, 
\end{equation}
where $\lambda\approx 0.22$ is the Cabibbo angle.  Clearly, an
additional suppression factor $\rho$ is required in $Y_U$ for those 
elements that contribute most significantly to the up and charm quark mass 
eigenvalues,
\begin{equation}\label{eq:yup}
Y_U \approx \left(\begin{array}{ccc}  0 & u_1\epsilon' \rho & 0 \\
-u_1\epsilon' \rho & u_2\epsilon \rho & u_3\epsilon \\
0 & u_4 \epsilon & u_5 \end{array}\right) \,\,\, ,
\label{eq:yu}
\end{equation}
where $u_1\ldots u_5$ are $O(1)$ coefficients. By embedding the U(2) model 
in a grand unified theory it is possible to obtain $\rho\approx \epsilon$ 
naturally; the model can then accommodate all the desired fermion mass 
hierarchies for choices of the coefficients $u_i$ and $d_i$ that 
are all of order one~\cite{u23}.  For example, in an SU(5) GUT,
$Y_U$ is associated with the coupling ${\bf 10}$-${\bf 10}$-${\bf 5}$,
where the ${\bf 10}$'s represent matter fields, and the ${\bf 5}$ is
the Higgs field $H$.  However, 
\begin{equation}
{\bf 10}\otimes{\bf 10} = {\bf \overline{5}}_s \oplus 
{\bf \overline{45}}_a \oplus {\bf \overline{50}}_s  
\,\,\, ,
\end{equation}
where the subscripts indicate symmetry or antisymmetry under 
interchange of the two ${\bf \overline{10}}$'s.   If we assume 
that the antisymmetric flavon $A$ is 
an SU(5) singlet, the product $A H$ is a ${\bf 5}_a$, and does not 
contribute to $Y_U$.  Similarly, if the flavon $S$ is a ${\bf 75}$ with 
a vev in the hypercharge direction in SU(5) space, then the part of  
$S H$ that contains the Higgs doublet field transforms as a 
${\bf 45}_s$, which again does not contribute to $Y_U$.   To obtain 
nonvanishing couplings of the right size in the upper two-by-two block of $Y_U$ 
one introduces a singlet flavon $\Sigma$ that transforms as an SU(5) adjoint.
The vev of $S$ implies that the breakings of both U(2) to U(1) and SU(5) to 
the standard model gauge group are associated with vevs of order $\epsilon$.   
Thus, it is natural to assume $\langle \Sigma \rangle \approx \epsilon$,
which provides exactly the desired value of $\rho$ in Eq.~(\ref{eq:yup}).
Moreover, the SU(5) assignments for $A$ and $S$ provide for a Georgi-Jarlskog
mechanism~\cite{GJ}, so that unified U(2) models successfully account for the 
charged lepton mass spectrum as well.

While the textures that follow from the simple two-step breaking
of a U(2) flavor symmetry are indeed minimal, the original symmetry
group is not. It is natural to ask whether there are small discrete
groups that work equally well as horizontal symmetries. It was shown 
in Ref.~\cite{acl} that the charged fermion Yukawa textures usually associated 
with U(2) models may be reproduced assuming the symmetry $G_f=T'\times Z_3$, 
and the breaking pattern
\begin{eqnarray} 
T' \otimes Z_3  \stackrel{\epsilon}{\longrightarrow} Z^D_{3} 
\stackrel{\epsilon'}
{\longrightarrow} nothing.
\end{eqnarray}
Here, $T'$ is the double tetrahedral group, a discrete subgroup of 
SU(2) corresponding to the symmetry of a regular tetrahedron.  The 
factor $Z^D_{3}$ is the diagonal subgroup of a $Z_3 \subset T'$ and 
the additional $Z_3$ factor (see Section~\ref{minimal}).  Since U(2) is 
isomorphic to SU(2)$\times$U(1), it is not surprising that our 
discrete symmetry is a product of a discrete subgroup of
SU(2) and a discrete subgroup of U(1).  Moreover, it was argued in
Ref.~\cite{acl} that this symmetry is {\em minimal} in the sense 
that
\begin{itemize}
\item  $T'$ is the smallest discrete subgroup of SU(2) (and in fact the
smallest group of any kind) with 1-, 2- and 3-dimensional representations
and the multiplication rule {\bf 2}$\otimes${\bf 2}$=${\bf 3}$\oplus${\bf 1}.  
These two ingredients are necessary to reproduce the successful U(2) 
textures.
\item  $Z_3$ is the smallest discrete subgroup of U(1) that allows $G_f$ 
to contain a subgroup forbidding all order $O(\epsilon')$ entries in 
the Yukawa textures.   
\end{itemize}
The latter statement applies to models in which $T'$ is a discrete gauge
symmetry (see Section~\ref{gauge}); models with a global $T'$ symmetry do not require 
any additional Abelian factors, as we demonstrate in Section~\ref{global}.  The use of 
discrete gauge rather than global symmetries is motivated by various 
arguments that the latter 
are violated at order one by quantum gravitational effects~\cite{worm}. 
In two of the models 
we present, $T'$ is an anomaly-free discrete gauge symmetry, while the 
additional $Z_n$ factor is not.  As in many of the Abelian models 
described in the literature~\cite{abelian}, 
we simply assume that the $Z_n$ factor may 
be embedded in a U(1) gauge symmetry whose anomalies are cancelled by 
the Green-Schwarz mechanism~\cite{GS}.  Thus, our models may be viewed 
as consistent low-energy effective theories for flavor symmetries that 
are local in a complete, high-energy theory.  

On a more practical level, the different representation structure of $T'$ allows for 
elegant solutions to the solar and atmospheric neutrino problems that do not
alter the predictive quark and charged lepton 
Yukawa textures, nor require the introduction of sterile neutrinos. While
similar results can be obtained  in some SO(10)$\times$U(2) models~\cite{barcre},
we obtain our successful solutions using a much smaller symmetry 
structure.\footnote{For a similar approach, see Ref.~\cite{raby}.}
One goal of this paper is to study 
these solutions at a level of quantitative detail not presented in 
our earlier work.

In addition, we propose two new models involving $T'$ symmetry. The first 
model, based on the discrete gauge symmetry $T'\times Z_6$, reproduces all 
important features of the SU(5)$\times$U(2) model {\em without requiring
a field-theoretic grand unified theory}.  In other words, 
the suppression of $m_u$ and $m_c$ in the SU(5)$\times$U(2) 
theory described earlier is achieved in $T'\times Z_6$ without SU(5).  
In addition, the ratio $m_b/m_t$, which is not explained in SU(5)$\times$U(2), 
is predicted in our model to be of $O(\epsilon) \approx 0.02$ for 
$\tan\beta\sim O(1)$, where $\tan\beta$  is the ratio of Higgs field vevs
$\langle H_U \rangle / \langle H_D \rangle$.  In a second model, we 
consider the implications of $T'$ as a purely global flavor symmetry.  
Although in this case the symmetry may not be fundamental, it could still
arise as an accidental symmetry at low energies.  We show that it 
is possible to construct a viable model based on $T'$ alone,
with no additional Abelian factors.  While it is well known that
supersymmetric models with a continuous SU(2) flavor symmetry and
a ${\bf 2} \oplus {\bf 1}$ representation structure
do not have viable Yukawa 
textures, our global $T'$ model demonstrates that discrete subgroups of
SU(2) remain viable alternatives.

Our paper is organized as follows.  In the next section, we
discuss the meaning of discrete gauge symmetries and the
relevant anomaly-cancellation constraints in the low-energy
effective theory.  In Sections~\ref{tprime} and~\ref{minimal}, 
we review the group theory of
$T'$ and present the minimal model described in Ref.~\cite{acl}.
In Section~\ref{numerical} we fit predictions of the model to charged fermion 
and neutrino masses and mixing angles, including the most significant 
renormalization group effects. In Section~\ref{su5u2}, 
we present the $T'\times Z_6$ 
model that reproduces the important features of the SU(5)$\times$U(2)
model with neither SU(5) nor U(2).  
In Section~\ref{global}, we show how to construct a viable global $T'$ model with no 
Abelian factors.  In Section~\ref{sterile} we comment on one scenario involving 
sterile neutrinos, and in the final section we summarize our conclusions.

\section{What Is a Discrete Gauge Symmetry?} \label{gauge}

Let us define a discrete gauge symmetry provisionally as a discrete 
remnant of a spontaneously broken continuous gauge symmetry.  Below
the breaking scale $\Lambda$ of the continuous symmetry, the 
low-energy effective Lagrangian has interactions that are invariant 
under the unbroken discrete group, no massless gauge fields,
and derivatives that transform trivially.
It would seem then that this effective theory is identical to one 
with a purely global discrete symmetry.  In this section, we
review the arguments suggesting that this is not the case. 
We first illustrate how gauge invariance of a theory spontaneously
broken to a discrete subgroup dictates the form of all terms in the
low-energy effective theory, and thus renders its discrete invariance
immune to wormhole dynamics. We then show that a
theory with a discrete gauge symmetry predicts topological defects
not present in a theory with a global symmetry, and that these play 
an important role in demonstrating that discrete gauge charges leave 
quantum-mechanical hair on black holes. Both observations suggest that 
discrete gauge symmetries are viable as candidates 
for fundamental symmetries of nature.  After reviewing these arguments we 
summarize the anomaly-cancellation constraints relevant to 
low-energy theories with discrete gauge symmetries. We use these
constraints in constructing models throughout this paper.

Following a discussion by Banks~\cite{banks}, let us consider the low-energy
effective theory that results from spontaneously breaking a U(1) 
gauge symmetry to a discrete subgroup.  The full theory consists
of two scalar fields $\chi$ and $\phi$ with U(1) charges $q$ and
$1$, respectively, where $q$ is an integer.  The Lagrangian is the 
usual one for an Abelian Higgs model:
\begin{equation}
{\cal L} = -\frac{1}{4g^2} F_{\mu\nu} F^{\mu\nu}
+ |\partial_\mu \chi - i q A_\mu \chi|^2
+ |\partial_\mu \phi - i A_\mu \phi|^2 + V(\chi^\dagger \chi).
\end{equation}
The potential $V$ is such that the $\chi$ field acquires a vacuum 
expectation value $\Lambda$.  Let us rewrite the Lagrangian
using the nonlinear field redefinition $\chi = (\Lambda + \sigma)
e^{i\theta}/\sqrt{2}$.  This yields
\begin{equation}\label{eq:siglag}
{\cal L} = 
-\frac{1}{4g^2} F_{\mu\nu} F^{\mu\nu} 
+\frac{1}{2} \partial_\mu \sigma \partial^\mu \sigma
+ \frac{1}{2} (\Lambda + \sigma)^2 (\partial_\mu\theta
-q A_\mu)^2 + |\partial_\mu \phi - i A_\mu \phi|^2 
+ V(\sigma) \,\,\, ,
\end{equation}
where $\sigma$ is the Higgs field and $\theta$ is the would-be 
longitudinal component of the U(1) gauge boson in unitary gauge. 
We choose to construct a low-energy effective theory in which the 
$\sigma$ field, which has a mass of order $\Lambda$, is integrated out. 
However, we retain the gauge field $A_\mu$ as well as the unphysical scalar
field $\theta$.  Although the gauge symmetry is spontaneously broken, the
Lagrangian of the theory remains invariant under the local U(1) 
transformation:
\begin{equation}
\phi \rightarrow e^{i \alpha(x)} \phi, \,\,\,\,\,\, 
A^\mu \rightarrow A^\mu + \partial_\mu \alpha(x), \,\,\,\,\,\, 
\theta \rightarrow \theta + q \,\alpha(x) .
\end{equation} 
The low-energy effective Lagrangian then consists of the kinetic terms
\begin{equation}\label{eq:kterms}
{\cal L}= -\frac{1}{4g^2} F_{\mu\nu} F^{\mu\nu}
+ |\partial_\mu \phi - i A_\mu \phi|^2 +
\frac{1}{2}\Lambda^2 (\partial_\mu \theta - q A_\mu)^2 \,\,\, ,
\end{equation}
as well as the most general set of gauge-invariant operators involving 
the fields $\phi$, $e^{i\theta}$, and covariant derivatives, with powers
of $\Lambda$ included to obtain the correct mass dimensions.  We can classify 
the interactions in the effective Lagrangian that involve $\phi$ into two types: 
terms that are invariant under global U(1) transformations on $\phi$ alone (with the 
other fields held fixed), and those that are not.  A typical term of the first 
type is $\phi^\dagger\phi$; terms of the second type necessarily involve the 
U(1) gauge-invariant product
\begin{equation}
e^{-i \theta} \phi^q \,\, ,
\end{equation}
or similar products with derivatives.  Such terms are invariant under
a $Z_q$ phase rotation of the field $\phi$ alone.   Thus, gauge invariance of
the low-energy theory implies it must have an unbroken $Z_q$ symmetry.  Since 
this is a consequence of a local symmetry, it cannot be violated by wormhole 
dynamics.

We now show that information on discrete gauge charges is not lost when a
charged particle falls into a black hole.  To do so, first note that the 
Abelian Higgs model has stable cosmic string solutions.  In the case where 
$\phi=0$, the kinetic energy terms in Eq.~(\ref{eq:kterms}) 
are minimized when 
\begin{equation}\label{eq:min}
A_\mu = \frac{1}{q}\partial_\mu \theta \,\, .
\end{equation}
For nonsingular gauge field configurations, this is related to
$A_\mu=0$ by a gauge transformation.  However, singular solutions
also exist;  a cosmic string along the $x^3$ axis corresponds to
\begin{equation}
A_i = \frac{1}{q}\epsilon_{ij} \frac{x^j}{x_1^2+x_2^2}\,\, , \,\,\,\,\, 
i,j = 1,2\,\, , \,\,\,\,\,\,\,\,\, \theta=\arctan(x_2/x_1) \,\,\, .
\end{equation}
If one couples the gauge field to a classical current $j^\mu$, then
the change in the action by adding one such cosmic string is
\begin{equation}
\delta S = (1/q) \int \partial_\mu \theta j^\mu\,\, ,   \,\,\,
\end{equation}
which follows from Eq.~(\ref{eq:min}).  Taking $j^\mu$ to
be the current of a particle with unit U(1) charge (and hence
nontrivial $Z_q$ charge) that circles the string, one finds 
that
\begin{equation}
\delta S = \frac{2\pi}{q} \,\,\, .
\end{equation}
This implies an observable Aharanov-Bohm effect in
the scattering of particles with discrete gauge charge off 
cosmic strings.  Krauss and Wilczek~\cite{KW} use this observation
to argue that the scattering of a cosmic string off a particle
with discrete gauge charge that is falling into a black hole is
insensitive to the point at which the particle crosses the
event horizon.  Thus, the discrete charge of the particle
is not lost, and the black hole grows quantum-mechanical
hair.

It is interesting to note that the discussion above may
be rephrased in unitary gauge by making the initial replacements
\begin{equation}
B_\mu = A_\mu - (1/q)\partial_\mu \theta , \,\,\,\,\, 
\mbox{ and } \,\,\,\,\, \Phi = e^{-i\theta/q}\phi\,\, , \,\,\, 
\end{equation}
in Eq.~(\ref{eq:siglag}), which then becomes
\begin{equation}
{\cal L} = -\frac{1}{4g^2} F_{\mu\nu}F^{\mu\nu} + \frac{1}{2}
\partial_\mu\sigma\partial^\mu \sigma
+ \frac{1}{2} (\Lambda+\sigma)^2 q^2 B^\mu B_\mu
+ | \partial_\mu \Phi - iB_\mu \Phi |^2 + V(\sigma) \,\,\, .
\end{equation}
Unlike the previous approach, all the fields above are gauge-invariant; one 
may integrate out $B_\mu$ and $\sigma$, and obtain all the possible $Z_q$-invariant 
interactions involving the light field $\Phi$.   This formulation of the low-energy 
theory is peculiar in that the periodicity of $\theta$ implies that
\begin{equation}
e^{2 n \pi i/q} \Phi \equiv \Phi, \,\, {\rm for \,\,  all \,\, integers}
\,\,  n.
\end{equation}
Thus, the field manifold of $\phi$ is not the complex plane $C$, 
but rather the orbifold $C/Z_q$: Field configurations connected by $Z_q$ 
transformations are identified, and hence are physically redundant, the
hallmark of a gauge symmetry. Given 
this manifold, the field $\Phi$ has a conical singularity at the origin in 
field space; strings in unitary gauge correspond to $\Phi$ field configurations 
that wrap around this singularity as the azimuthal angle varies 
from $0$ to $2\pi$.  

As the previous U(1) $\rightarrow Z_q$ example demonstrates, a discrete
gauge symmetry can arise in a renormalizable field theory when a
continuous gauge symmetry is spontaneously broken by a Higgs field vev
that leaves a discrete symmetry unbroken. The same can occur for non-Abelian
symmetries as well. For example, one may break a gauged SU(2) symmetry
with a Higgs field transforming as a ${\bf 7}$ (which contains a $T'$ singlet),
leaving the theory invariant under $T'$. On the other hand, the U(1) $\rightarrow 
Z_q$ example suggests how a discrete symmetry may be defined without an 
explicit embedding in a continuous group. In string theory, the discrete
symmetry may be a remnant of general coordinate invariance, ordinary
gauge invariance, or the larger gauge symmetry of string theory~\cite{banks}.
For our purposes, however, the nature of the high energy theory is irrelevant.

It is worth mentioning in passing that spontaneously-broken discrete gauge 
symmetries have domain walls that are not topologically stable. Holes bounded by 
strings may spontaneously nucleate, allowing the walls to tear themselves to pieces 
while dissipating energy through gravitational radiation~\cite{houches}.
The effectiveness of this 
mechanism at avoiding cosmological problems is not relevant to our discussion 
since the flavor-symmetry-breaking scale in our models is high enough (of order the 
unification scale) that all topological defects are eliminated by inflation.

Finally, it is relevant to consider whether there are any constraints on 
the low-energy particle content of theories with discrete gauge symmetries.
Since continuous gauge symmetries must satisfy anomaly-cancellation
conditions, the particle content of low-energy theories with discrete 
gauge symmetries is restricted.  Ib\'{a}\~{n}ez and Ross~\cite{IR} were the first
to consider the constraints on a discrete gauged $Z_q$ symmetry, and their results 
were refined by Banks and Dine~\cite{BD}:  Let $G_0$ be a simple factor of
the continuous group in which
a discrete gauge symmetry is embedded, and let $G_A$ and $G_N$ represent
the unbroken Abelian and non-Abelian gauge symmetries of the low-energy effective
theory.  Cancellation of the $G_0 G_N^2$ anomaly is the only new requirement
for consistency of the low-energy theory; all other anomaly-cancellation
constraints involving $G_0$ can be satisfied by the introduction of heavy states. 
Banks and Dine point out that this requirement, termed the linear
Ib\'{a}\~{n}ez-Ross condition, is equivalent to demanding discrete gauge 
invariance of nonperturbative interactions generated by instantons of the 
unbroken continuous gauge groups.  This observation demonstrates that
consistency of a discrete gauge symmetry at low energies can be established 
without reference to any particular embedding.

\section{The Group $T^\prime$} \label{tprime}

All of the symmetries described in this paper contain $T^\prime$, the
double tetrahedral group.\footnote{ For a review of basic terms of
discrete group theory, see Ref.~\cite{d6}, Appendix A.} Geometrically, 
$T^\prime$ is
defined as the group of all 24 proper rotations in three dimensions
leaving a regular tetrahedron invariant in the SU(2) double covering
of SO(3).  This perhaps opaque definition may be understood in the
following way.  There exists a group of 12 elements called the
tetrahedral group $T$, consisting of all proper rotations in three
dimensions leaving a regular tetrahedron invariant (Fig.~1).  It is
constructed by parameterizing the group SO(3) of {\it all\/} proper
rotations in three dimensions in terms of familiar Euler angles, and
then restricting to those discrete values of angles describing
rotations taking a regular tetrahedron into coincidence with itself.
The same Euler angles describe rotations in SU(2) space, since SU(2)
and SO(3) are locally isomorphic, so that $T^\prime$ is the subgroup of
SU(2) corresponding to the same Euler angles as $T \subset$ SO(3).
One therefore expects that even-dimensional representations of
$T^\prime$ are spinorial, i.e., are multiplied by $-1$ under a $2\pi$
rotation (called $R$ in the literature), while odd-dimensional
representations of $T^\prime$ coincide with those of $T$ and are
invariant under this rotation, as may be verified by the character
table,
Table~\ref{char}.

$T^\prime$ is generated by the rotations $C_2$ and $C_3$ depicted in
Fig.~1.  Because of the double-valued nature of $T^\prime$ rotations, 
these elements actually have orders 4 and 6, respectively.
For reasons to be described below, it turns out to be convenient to
present explicit representations ({\it reps}) for an 
element of order 4 (such as
$C_2$) and one of order 3 (such as $C_3 R$).  We label these elements
as $g_5$ and $g_9$, respectively;\footnote{The element labels are
  chosen to coincide with those of Thomas and Wood, Ref.~\cite{TW},
  where $T^\prime$ is seen to be isomorphic to SL$_2$(F$_3$), the
  group of two-by-two unimodular matrices whose elements are added
  and multiplied as integers modulo 3.} then $T^\prime$ is defined
by the multiplication rules $g_9^3 = g_5^4 = 1$, $g_9 g_5^2 = g_5^2
g_9$, and $g_5 g_9^{-1} g_5 = g_9 g_5 g_9$.  One may then show that
each of the 24 elements may be written uniquely in the canonical form
$g_9^p g_5^q g_9^r$, where $p = 0, \pm 1$, and if $q = 0$ or 2 then
$r = 0$, while if $q = \pm 1$ then $r = 0, \pm 1$.

The group $T'$ is central to our model building since it is the smallest
with 1-, 2-, and 3-dimensional reps and the multiplication rule 
${\bf 2} \otimes {\bf 2} = {\bf 3} \oplus {\bf 1}$. $T'$ models
therefore allow for flavons that perform the same roles as $\phi_a$,
$S_{ab}$ and $A_{ab}$ in the U(2) model.  The only other 
24-element group that has reps of the same dimensions
is the octahedral group $O$ (which is isomorphic to $S_4$).  
In this case, however, the product of two doublet reps does not 
contain a triplet, and the analogy to U(2) is lost.

More specifically, $T'$ has three singlets ${\bf 1}^0$ and ${\bf 1}^{\pm}$, 
three doublets, ${\bf 2}^0$ and ${\bf 2}^{\pm}$, and one triplet, ${\bf 3}$.
The {\em triality} superscript provides a concise way of stating the 
multiplication rules for these reps: With the identification of $\pm$ 
as $\pm 1$, trialities add under addition modulo three, and the following
rules hold:
\begin{equation}
\begin{array}{lcl}
{\bf 1} \otimes {\bf R} = {\bf R} \otimes {\bf 1}  =  {\bf R}
\hspace{1em}\mbox{\rm for any rep ${\bf R}$},  &\hspace{3em} &
{\bf 2} \otimes {\bf 2}  =  {\bf 3} \oplus {\bf 1}, \\
{\bf 2} \otimes {\bf 3} = {\bf 3} \otimes {\bf 2}  =  {\bf 2}^0
\oplus {\bf 2}^+ \oplus {\bf 2}^- , &\hspace{3em} &
{\bf 3} \otimes {\bf 3}  =  {\bf 3} \oplus {\bf 3} \oplus {\bf 1}^0
\oplus {\bf 1}^+ \oplus {\bf 1}^- . \end{array}
\end{equation}
Note that trialities flip sign under Hermitian conjugation, so that
${\bf 2}^+ \otimes {\bf 2}^- = {\bf 3} \oplus {\bf 1}^0$ while 
$({\bf 2}^+)^\dagger \otimes {\bf 2}^- = {\bf 3} \oplus {\bf 1}^+$.

The multiplication of $T'$ representations may be made explicit
by the use of Clebsch-Gordan matrices. For example, let the fields 
$\chi$ and $\psi$ be column vectors that transform as ${\bf 2^+}$ and
${\bf 2^-}$ under $T'$, respectively.  From the rules above, we know 
that the product of these reps contains a trivial singlet, the ${\bf 1}^0$,
but it is not immediately clear how to construct this representation
out of the given fields.  Formally, we seek a matrix
$M$ such that the product
\begin{equation}\label{eq:examp}
\chi^T M \psi \rightarrow \chi^T M \psi
\end{equation}  
under the transformations $\chi \rightarrow R(g) \chi$ and
$\psi \rightarrow R(g) \chi$, where $R$ is a two-dimensional matrix 
rep, and $g$ runs over all elements of the group.  From our earlier
discussion, it is only necessary that we consider transformations
associated with the defining elements, $g_5$ and $g_9$, to solve for 
the form of $M$; in the present case, one finds that $M$ is proportional 
to the Pauli matrix $\sigma_2$.  This algebraic procedure is easily 
generalized to products of other reps.  Explicit matrix representations for 
the generating elements $g_5$ and $g_9$, as well as the complete
set of Clebsch-Gordan matrices for combining $T'$ reps are provided
in Appendix~A.  The reader should keep in mind that these Clebsch-Gordan 
matrices must be taken into account if one is to reproduce
the Yukawa textures presented later in this paper.  For example,
without the factor of $\sigma_2$, one might not realize that a vev in 
the first component of $\chi$ couples only to the second 
component of $\psi$.

As mentioned in the Introduction, we also require that our discrete flavor 
symmetry contain a subgroup that rotates first-generation matter fields by a 
phase.  This subgroup plays the same role as the intermediate U(1) symmetry in the 
U(2) model, and must forbid all entries in the first row and column of each Yukawa 
matrix.   The smallest discrete subgroup that one might consider is a $Z_2$ 
that flips the sign of all first generation matter fields.  Unfortunately, such a 
transformation leaves the 11 entry of each Yukawa matrix invariant (two
sign flips), so that the 
up and down quarks could, in principle, acquire masses that are too large. A $Z_3$ 
phase rotation, on the other hand, does not lead to the same problem, and a $Z_3$ subgroup 
of $T'$ is generated by the element $g_9$ defined previously.  From Appendix~A, 
we see that the two-dimensional representation matrices for the element $g_9$ 
are given by
\begin{equation}
g_9({\bf 2}^0) = \left(\begin{array}{cc} \eta^2 & 0 \\ 0 & \eta \end{array}\right), \,\,\,\,\,
g_9({\bf 2}^+)= \left(\begin{array}{cc} 1 & 0 \\ 0 & \eta^2 \end{array}\right), \,\,\,\,\, 
g_9({\bf 2}^-)= \left(\begin{array}{cc} \eta & 0 \\ 0 & 1 \end{array}\right) ,
\end{equation}
where $\eta \equiv \exp(2\pi i/3)$.  If matter fields of the first two generations
are assigned to the ${\bf 2}^-$ rep, one then obtains the desired phase rotation under
the $Z_3$ subgroup.  This observation is at the heart of the global $T'$ model
presented in Section~\ref{global}.\footnote{One can also imagine models in which
the symmetry group breaks to a non-Abelian subgroup; however, in this case the
simple rephasing of multiplet components under the subgroup is not guaranteed.}  

As we see below, however, models in which $T'$ is free of discrete gauge 
anomalies are much easier to construct if matter fields are assigned to the ${\bf 2}^0$ 
rep instead.   In this case, let us consider extending the flavor symmetry group to 
$T'\times Z_3$.   We identify a new triality index $0$, $+$ and $-$ with the $Z_3$ phase 
rotations $1$, $\eta$, and $\eta^2$, respectively.  Like the $T'$ indices, the $Z_3$ 
trialities also combine via addition modulo 3.  Reps of $T'\times Z_3$ are denoted by 
affixing this additional triality as a superscript, e.g., ${\bf 2}^{+-}$.  
We now identify the desired intermediate symmetry as the diagonal subgroup 
of the original $Z_3$, generated by the element $g_9$, and the new $Z_3$ factor.  
We call this subgroup $Z_3^D$ henceforth.  It is easy to see that the rep ${\bf 2}^{0-}$ 
transforms under $Z_3^D$ by the matrix 
\begin{equation}
\left(\begin{array}{cc} \eta & 0 \\ 0 & 1 \end{array}\right) 
\end{equation}
which is simply the product of $g_9({\bf 2}^0)$ and $\eta^2$.  The matter field 
assignments ${\bf 2}^{0-} \oplus {\bf 1}^{00}$, and the breaking pattern $T'\times Z_3 \rightarrow Z_3^D \rightarrow
nothing$ are at the heart of the minimal flavor model discussed in the next 
section.  It is worth pointing out that the reps ${\bf 1}^{00}$, ${\bf 1}^{+-}$,
${\bf 1}^{-+}$, ${\bf 2}^{0-}$, ${\bf 2}^{++}$ and ${\bf 2}^{-0}$ are special
in that these singlet reps and the second component of the doublets remain
invariant under $Z_3^D$.  Thus any ${\bf 2} \oplus {\bf 1}$ combination
of these reps is potentially useful in building models with U(2)-like textures.

Finally, we return to the issue of anomaly cancellation.  We pointed
out in Section~\ref{gauge} that consistency of a 
discrete gauge symmetry at low energies
only requires the cancellation of anomalies that (1) involve the unbroken
non-Abelian continuous gauge groups, and (2) are linear in a continuous
group in which the discrete group is embedded.  If we embed $T'$ in SU(2),
then these constraints are satisfied {\em automatically}, providing
that the particle content of a given model fills complete SU(2) representations.  
Let us therefore consider the embedding of $T'$ in SU(2) in more detail.

The group SU(2) has one rep of each nonnegative integral dimension $n$ 
(the spin $(n-1)/2$ rep), while $T^\prime$ has only singlet, doublet, and 
triplet reps.  It must be the case that large SU(2) reps break up into a 
number of $T^\prime$ reps with the same total dimension.  To see this decomposition,
consider the characteristic polynomial of matrices in each of the $T^\prime$ reps 
for any two rotations that generate the full group.  The same can be done for the
full SU(2) group restricted to the particular Euler angles that give
$T^\prime$.  Then a large rep matrix of SU(2) is block-diagonalizable
into smaller blocks corresponding to rep matrices of $T^\prime$; in
particular, the characteristic polynomial of the SU(2) matrix is the
product of those of the $T^\prime$ matrices.  It is then possible to
extract which $T^\prime$ reps appear in a given SU(2) rep, as well as
their multiplicities.  The results of this decomposition are summarized in
Table~\ref{decomp}.  There we see that the ${\bf 1}^0$, ${\bf 2}^0$,
and ${\bf 3}$ reps of $T^\prime$ correspond to the complete ${\bf 1}$,
${\bf 2}$, and ${\bf 3}$ reps of SU(2).  It follows, for example, that $T'$ 
is non-anomalous in all models utilizing the ${\bf 2}^{0-} \oplus {\bf 1}^{00}$ 
representation structure for the matter fields (with Higgs fields as singlets).   
Note that there is no meaningful low-energy constraint on the $Z_3$
charges since Abelian factors may be embedded at high energies 
in U(1) gauge groups whose anomalies are cancelled by the Green-Schwarz 
mechanism~\cite{GS}.

\section{A Minimal Model} \label{minimal}
In this section we review the minimal $T' \times Z_3$ model presented in
Ref.~\cite{acl}, which we study in quantitative detail in 
Section~\ref{numerical}. The three generations of matter fields are 
assigned to the $T' \times Z_3$ reps ${\bf 2}^{0-} \oplus {\bf 1}^{00}$ 
while the Higgs fields $H_{U,D}$ are taken to be pure $G_f$ singlets. 
Given these assignments, it is easy to obtain the 
transformation properties of the Yukawa matrices,
\begin{eqnarray} \label{yuk}
Y_{U,D,L} & \sim & \left( \begin{array}{c|c} 
[ {\bf 3}^{-}\oplus {\bf 1}^{0-}] 
& [ {\bf 2}^{0+}] \\ 
\hline
{[} {\bf 2}^{0+}]
 & [ {\bf 1}^{00} ]
\end{array} \right) .
\end{eqnarray}
Equation~(\ref{yuk}) indicates the flavon reps 
needed to construct the fermion mass matrices, namely, ${\bf 1}^{0-}$,
${\bf 2}^{0+}$, and  ${\bf 3}^{-}$, which we call $A$, $\phi$, and
$S$, respectively.  Once these flavons acquire vevs, the flavor group 
is broken. We are interested in a two-step breaking
\begin{eqnarray} \label{break}
T' \otimes Z_3  \stackrel{\epsilon}{\longrightarrow} Z^D_{3} 
\stackrel{\epsilon'}
{\longrightarrow} nothing,
\end{eqnarray}
where $\epsilon' < \epsilon$ again represent ratios of flavon vevs to the
scale $M_f$.
Since we have chosen a `special' doublet rep for the first two generations,
which transforms as {\it diag}$\{\eta,1\}$ under $Z_3^D$, only the 22, 23, and
32 entries of the Yukawa matrices may develop vevs of $O(\epsilon)$
originating from vevs in $S$ and $\phi$.  The symmetry $Z_3^D$
is then broken by a ${\bf 1}^{0-}$ vev of $O(\epsilon')$.  The 
Clebsch-Gordan coefficient that couples a ${\bf 1}^{0-}$ to two
${\bf 2}^{0-}$ doublets is proportional to $\sigma_2$, so the $\epsilon'$
appears in an antisymmetric matrix. These considerations yield the textures
\begin{eqnarray} \label{textures}
Y_{U,D,L} & \sim & \left( \begin{array}{ccc} 
0 & \epsilon' & 0 \\
-\epsilon' & \epsilon & \epsilon \\ 
0 & \epsilon & 1
\end{array} \right) \,\, ,
\end{eqnarray}
where $O(1)$ coefficients have been omitted. Since the ${\bf 1}^{0-}$ 
and ${\bf 3}^-$
flavon vevs appear as antisymmetric and symmetric matrices,
respectively, all features of the grand unified extension of the U(2)
model are obtained here, assuming the same GUT transformation properties
are assigned to $\phi$, $S$, and $A$.  One can also show readily that 
the squark and slepton mass-squared matrices are the same as in the U(2) 
model.

It was shown in Ref.~\cite{acl} that this simple model can be extended to describe
the observed deficit of solar and atmospheric
neutrinos. Models for lepton masses were constructed both with and without the
assumption of SU(5) unification. The latter possibility is of interest,
for example, if one is only concerned with explaining flavor physics of
the lepton sector, and is provided for completeness. In either case, the
proposed extensions yield viable neutrino textures with naturally large
mixing between the second and third generations. Moreover, these extensions
do not alter the charged fermion textures, so that all the relations between
masses and mixing angles in the U(2) model are also predictions of
$T' \times Z_3$. We now review the two cases considered in Ref.~\cite{acl}.

Case I: Here we do not assume grand unification, so that all flavons
are SU(5) singlets.
We introduce three generations of right-handed neutrinos transforming as
\begin{eqnarray} \label{right}
\nu_R \sim {\bf 2}^{0-} \oplus {\bf 1}^{-+} .
\end{eqnarray}
Note that this representation choice differs from that of the other 
matter fields only in the third generation. Since $\nu_{R}$ are singlets
under the standard model gauge groups, introducing a ${\bf 1}^{-}$ field
by itself creates no anomaly problems.
The neutrino Dirac 
and Majorana mass matrices then allow flavons that do not contribute to the 
charged fermion mass matrices. Their transformation properties are
given by
\begin{equation}
M_{LR} \sim  \left( \begin{array}{c|c} [ {\bf 3}^{-} \oplus {\bf 1}^{0-}] 
& [ {\bf 2}^{+0}] \\ 
\hline
{[} {\bf 2}^{0+}]
 & [ {\bf 1}^{+-} ]
\end{array} \right) \,\,\, , \hspace{2em} 
M_{RR} \sim \left( \begin{array}{c|c} [ {\bf 3}^{-} ] 
& [ {\bf 2}^{+0}] \\ 
\hline
{[} {\bf 2}^{+0}]
 & [ {\bf 1}^{-+} ]
\end{array} \right) \,\,\, .
\end{equation}
Note that one obtains the same triplet and nontrivial singlet in the upper 
two-by-two block as in the charged fermion mass matrices, as well as
one of the same flavon doublets, the ${\bf 2}^{0+}$; the rep
${\bf 1}^{0-}$ is not present in $M_{RR}$, since Majorana mass matrices
are symmetric. In addition we obtain the reps 
${\bf 2}^{+0}$, ${\bf 1}^{+-}$, and ${\bf 1}^{-+}$, which did not appear 
in Eq.~(\ref{yuk}). New flavon fields can now be introduced with these 
transformation properties, and their effects on the neutrino physics 
explored. Let us introduce a single\footnote{Assuming
more than one $\phi_\nu$ leads to the same qualitative results.} new 
flavon $ \phi_{\nu}$ transforming as a ${\bf 2^{+0}}$ and with 
a vev
\begin{eqnarray} \label{phinu}
\frac{\langle {\bf \phi_{\nu}} \rangle}{M_f} \sim \sigma_2
\left( \begin{array}{c} \epsilon' \\
\epsilon \end{array} \right) \,\,\, ,
\label{eq:newvev}
\end{eqnarray}
where $\sigma_2$ is the Clebsch that couples the two doublets to 
${\bf 1}^{0-}$. This new flavon is the only extension 
we make to the model in order to describe the neutrino phenomenology. After 
introducing $\phi_\nu$, the neutrino Dirac and Majorana mass matrices read
\begin{equation} \label{MLRMRR}
M_{LR}  \approx  \left( \begin{array}{ccc} 0 
& l_1 \epsilon' & l_3 r_2 \epsilon' \\ -l_1
\epsilon' & l_2 \epsilon & l_3 r_1 \epsilon \\ 0 &
l_4 \epsilon & 0 \end{array} \right) \langle H_U \rangle \,\,\, ,
\hspace{2em}
M_{RR}  \approx  \left( \begin{array}{ccc} r_4 r_2 {\epsilon'}^{2} 
& r_4 r_1 \epsilon \epsilon' & r_2 \epsilon' \\ r_4 r_1
\epsilon \epsilon' & r_3 \epsilon & r_1 \epsilon \\ r_2 \epsilon' &
r_1 \epsilon & 0 \end{array} \right) \Lambda_R \,\,\, ,
\end{equation}
where $\Lambda_R$ is the right-handed neutrino mass scale, and we have
parameterized the $O(1)$ coefficients.  Furthermore, the charged lepton
Yukawa matrix including $O(1)$ coefficients reads
\begin{eqnarray} \label{YL}
Y_{L}  \approx  \left( \begin{array}{ccc} 0 & c_1 \epsilon' & 0 
 \\ -c_1 \epsilon' & 3c_2\epsilon & c_3\epsilon 

 \\ 0 & c_4\epsilon & c_5
\end{array} \right)\xi \,\,\, .
\end{eqnarray}
The factor of $3$ in the 22 entry is simply assumed at present, but 
originates from the Georgi-Jarlskog mechanism in the
grand unified case considered next.

The left-handed Majorana mass matrix $M_{LL}$ follows from the 
seesaw mechanism  
\begin{eqnarray} \label{seesaw}
M_{LL} \approx M_{LR} M_{RR}^{-1} M_{LR}^{T} \,\,\, ,
\end{eqnarray}
which yields
\begin{eqnarray} \label{MLL}
M_{LL}  \sim  \left( \begin{array}{ccc} (\epsilon'/\epsilon)^2 & 
\epsilon'/\epsilon & \epsilon'/\epsilon 
\\ \epsilon'/\epsilon & 1 & 1 
\\ \epsilon'/\epsilon & 1 & 1
\end{array} \right) \frac{\langle H_U \rangle^2 \epsilon}{\Lambda_R} ,
\end{eqnarray}
where we have suppressed the $O(1)$ coefficients. It is clear by inspection
that we naturally obtain large mixing between second- and third-generation
neutrinos. It is also important to point out that the two eigenvalues of
Eq.~(\ref{MLL}) that appear to be of $O(1)$ depend sensitively on the products
of a large number of order one coefficients. It is easy to obtain a
hierarchy of order $10$ in the two largest mass eigenvalues, without
allowing any of the coefficients defined in Eqs.~(\ref{MLRMRR})$-$(\ref{YL}) to
deviate from unity by more than a factor of 2. This comment is important in
understanding how the reasonable coefficient choices given in Ref.~\cite{acl}
account for the differing mass scales associated with atmospheric and
solar neutrino oscillations.

In order to determine neutrino oscillation parameters precisely one
needs to compute the neutrino CKM matrix.
If $M_{LL}$ and $Y_{L}$ are diagonalized by 
$M_{LL} = W M_{LL}^{0} W^{\dagger}$,
$Y_{L} = U_{L} Y_{L}^{0} U_{R}^{\dagger}$, then
\begin{eqnarray} \label{CKM}
V = U_{L}^{\dagger} W.
\end{eqnarray}
We parameterize this matrix as in Ref.~\cite{barhall},
\begin{eqnarray} \label{ckm}
V = \left(\begin{array}{ccc} c_{12} c_{13} & c_{13} s_{12} & s_{13} \\
-c_{23} s_{12} e^{i\phi} -c_{12} s_{13} s_{23} & c_{12} c_{23} e^{i\phi} 
-s_{12} s_{13} s_{23} & c_{13} s_{23} \\
s_{23} s_{12} e^{i\phi} -c_{12} c_{23} s_{13} & -c_{12} s_{23} e^{i\phi} 
-c_{23} s_{12} s_{13} & c_{13} c_{23}
\end{array} \right) \,\, ,
\end{eqnarray}
where $c_{ij} (s_{ij})$ 
stands for $\cos \theta_{ij} (\sin \theta_{ij})$. Then one finds
\begin{eqnarray} \label{sins}
\sin^{2}(2 \theta_{12}) = 4 \frac{V_{11}^{2} V_{12}^{2}}{(V_{11}^{2} + 
V_{12}^{2})^{2}} \,\, , \\
\sin^{2}(2 \theta_{23}) = 4 \frac{V_{23}^{2} V_{33}^{2}}{(V_{23}^{2} + 
V_{33}^{2})^{2}} \,\, .
\end{eqnarray}

The observed atmospheric neutrino fluxes may be explained by 
$\nu_\mu$-$\nu_\tau$ mixing
if $\sin^2 2\theta_{23} \agt 0.8$ and $10^{-3}\alt \Delta m^2_{23} \alt 
10^{-2}$, while the solar neutrino deficit may be accommodated 
by $\nu_e - \nu_{\mu}$ mixing assuming the 
small-angle MSW solution $2\times 10^{-3} \alt \sin^2 2\theta_{12} 
\alt 10^{-2}$ for $4\times10^{-6} \alt \Delta m^2_{12} \alt 10^{-5}$, where 
all squared masses are given in eV$^2$~\cite{Kam,bah}. 
These regions of parameter
space are the ones obtained most naturally from our models.\footnote{
The experimental ranges for neutrino mixing parameters follow from
a two-neutrino mixing approximation which is valid only if the mixing
angle $\theta_{13} < 15^{\circ}$~\cite{barhall}. This condition is satisfied
in all our models.}
Since $\Lambda_R$ is not determined
from symmetry considerations, it is only necessary to reproduce
$\Delta m_{23}^2 / \Delta m_{12}^2$. In Ref.~\cite{acl} a choice for
the $O(1)$ coefficients can be found that yields neutrino
mass ratios and mixing angles falling within the desired ranges
given above.
   
Case II: Here we assume SU(5) unification and that the flavons 
transform nontrivially under the GUT group, 
namely, $A \sim {\bf 1}$, $S \sim {\bf 75}$,
$\phi \sim {\bf 1}$, and $\Sigma \sim{\bf 24}$. Note that since 
$\overline{H} \sim {\bf \overline{5}}$, the products $S\overline{H}$ 
and $A \overline{H}$ transform as a ${\bf \overline{45}}$ 
and ${\bf \overline{5}}$, respectively, ultimately providing a factor 
of $3$ enhancement in the 22 entry of $Y_{L}$ (the Georgi-Jarlskog 
mechanism). In addition, two ${\bf 2}^{+0}$ doublets are 
introduced, $\phi_{\nu 1}$ and $\phi_{\nu 2}$, since the texture 
obtained for the neutrino masses by adding only one 
extra doublet is not viable.  Both doublets $\phi_\nu$ have vevs of 
the form displayed in Eq.~(\ref{phinu}).  As before, the presence of
these two new doublets does not alter the form of any charged fermion
Yukawa texture.

The neutrino Dirac and Majorana mass matrices now take the forms
\begin{equation} \label{mlrmrrgut}
M_{LR}  \approx \left( \begin{array}{ccc} 0 
& l_1 \epsilon' & l_5 r_2 \epsilon' \\ -l_1
\epsilon' & l_2 \epsilon^{2} & l_3 r_1 \epsilon \\ 0 &
l_4 \epsilon & 0 \end{array} \right) \langle H_U \rangle \,\,\, ,
\hspace{2em}
M_{RR}  \approx  \left( \begin{array}{ccc} r_3 {\epsilon'}^{2} 
& r_4 \epsilon \epsilon' & r_2 \epsilon' \\ r_4
\epsilon \epsilon' & r_5 \epsilon^{2} & r_1 \epsilon \\ r_2 \epsilon' &
r_1 \epsilon & 0 \end{array} \right) \Lambda_R \,\,\, ,
\end{equation}
while the charged lepton mass matrix is the same as in Eq.~(\ref{YL}). 
Using Eq.~(\ref{seesaw}) one obtains the texture:
\begin{eqnarray} \label{MLLGUT}
M_{LL}  \sim  \left( \begin{array}{ccc} (\epsilon'/\epsilon)^{2} & 
\epsilon{'}/\epsilon & \epsilon{'}/\epsilon 
\\ \epsilon{'}/\epsilon & 1 & 1 
\\ \epsilon{'}/\epsilon & 1 & 1
\end{array} \right) \frac{\langle H_U \rangle^2}{\Lambda_R} \,\,\, .
\end{eqnarray}
Again, a viable set of $O(1)$ coefficients may be found in Ref.~\cite{acl}.

While the texture in Eq.~(\ref{MLLGUT}) appears to be the same as the one
in Eq.~(\ref{MLL}) (up to an overall factor of $\epsilon$), there is in 
fact an important difference: the $O(1)$ entries in Eq.~(\ref{MLLGUT})
have a vanishing determinant at lowest order. The ratio of the two largest
eigenvalues are therefore determined by higher order corrections, which
must be taken into account to obtain the correct numerical results.\footnote{
In fact, the analysis made for 
the model in Case I included higher order terms, which did 
not contribute in any significant way.} While the zero determinant is
lifted at $O(\epsilon)$ in the superpotential, it is interesting that, in
this particular case, a comparable correction comes from D-terms that
alter the canonical form of the neutrino kinetic energy
\begin{equation} \label{kinetic}
\int d^{4} \theta [ \nu_{L}^{\dagger}\nu_{L} + 
\nu_{L}^{\dagger} B \nu_{L} ] \,\, .
\end{equation}
Here $B$ is a Hermitian matrix that depends on the flavons in the
model. The kinetic terms may be put back into canonical form by the
superfield redefinition $ \nu_{L} \rightarrow \sqrt{1 - B} \nu_{L}
\approx (1 - B/2) \nu_{L}$. This in turn leads to a correction to
$M_{LL}$,
\begin{eqnarray} \label{extramll}
M_{LL} \rightarrow M_{LL} -1/2  \{ B , M_{LL} \}.
\end{eqnarray} 
Numerically, it is only necessary that we retain the largest elements
of $B$
\begin{equation} \label{B}
B \approx \left( \begin{array}{ccc} \cdot & \cdot & \cdot \\
\cdot & \cdot & a \epsilon \\ \cdot & a \epsilon & \cdot 
\end{array} \right),
\end{equation} 
which also leads to an $O(\epsilon)$ correction to the determinant
discussed above. The parameter $a$ is included in the quantitative
analysis of the model presented in the next section.

\section{Numerical Analysis} \label{numerical}

The numerical check of the unified $T' \times Z_3$ model
presented in \cite{acl} relied on two
assumptions.  The first is that there exist $O(1)$ coefficients $c_i$,
$d_i$, and $u_i$ for the charged fermion Yukawa matrices that, when
combined with the particular choice of neutrino Yukawa parameters
$l_i$ and $r_i$, yield charged fermion mass eigenvalues and mixing
angles in agreement with the values observed.  This should not be a
problem since the textures of the $T^\prime \times Z_3$ model for the
charged fermions agree completely with those of the
U(2) model\cite{u23}, in which all of these observables are accommodated
in detailed fits. Second, the textures as written in the last section
are defined at the
scale $M_{\rm GUT} \approx 2 \times 10^{16}$ GeV, 
while the observables are of course measured below the
electroweak scale.  A truly meaningful fit requires running the gauge
and Yukawa couplings over this range. While the textures renormalized at
$M_{\rm GUT}$ and $m_t$ should not differ wildly in form, a
global fit is required to properly compare the predictions of our 
model to the experimental data. The purpose of this section is to report
on the necessary steps in these fits and the numerical results.

	In order to study the renormalization of gauge and Yukawa
couplings, we run the one-loop renormalization group equations (RGE's)
of the MSSM\cite{BBO} from $M_{\rm GUT}$  down to the electroweak scale 
taken to be $m_t = 175$ GeV.  This analysis 
does not include two-loop corrections nor
threshold effects at either end of the spectrum. In particular, 
this approach does not differentiate between the scales
$M_f$, $\epsilon M_f \approx M_{\rm GUT}$, $\epsilon^\prime M_f
\approx \epsilon'M_{\rm GUT}/\epsilon$, and $\Lambda_R \approx
\epsilon M_{\rm GUT}$.\footnote{Notice that $\Lambda_R \approx 
\epsilon M_{\rm GUT}$ yields the appropriate mass scale in 
Eq.~(\ref{MLLGUT}) for atmospheric neutrino oscillations.}
In any case, both the two-loop and threshold effects are
formally of subleading order, and therefore are taken into account
by permitting theoretical uncertainties in the gauge 
and Yukawa couplings of $O(1/16\pi^2)
\approx 1\%$.

	Values of the gauge couplings at $M_{\rm GUT}$ are obtained by
starting with the precision values extracted at the scale $M_Z$
~\cite{PDG},
\begin{eqnarray}
\alpha_1^{-1} (M_Z) & = & 59.99  \pm 0.04,   \nonumber \\
\alpha_2^{-1} (M_Z) & = & 29.57  \pm 0.03,   \nonumber \\
\alpha_3^{-1} (M_Z) & = & 8.40   \pm 0.13.
\end{eqnarray}
The gauge couplings are run from $M_Z$ to $m_t$ using the one-loop
Standard Model (SM) RGE's, and then from $m_t$ to $M_{\rm GUT}$ using the
one-loop MSSM RGE's.\footnote{It should be pointed out that, 
while the SM RGE's make use of
the $\overline{MS}$ scheme, the MSSM RGE's in Ref.~\cite{BBO} make use
of the $\overline{DR}$ scheme~\cite{AKL}, which differ at the matching
scale ($m_t$ by our choice) by an amount
\begin{equation}
\frac{4\pi}{\alpha^{\overline{DR}}_i} =
\frac{4\pi}{\alpha^{\overline{MS}}_i} - \frac 1 3 (C_A)_i,
\end{equation}
where $C_A = \{ 0, 2, 3 \}$ for $i=1,2,3$.}
 The GUT scale couplings are taken directly
from the textures of Eqs.~(\ref{eq:ydown}),~(\ref{eq:yup}),~(\ref{YL}),
and~(\ref{mlrmrrgut}), given numerical values for the 
dimensionless coefficients $c_i$, $d_i$, $l_i$, $r_i$, $u_i$, 
and $a$ (collectively $k_i$), and for $\epsilon$, 
$\epsilon'$, $\rho$, and $\xi$. The Yukawa 
matrices are then run down to $m_t$ and diagonalized.\footnote{
The RGE's are integrated by means of the Runge-Kutta method
with adaptive stepsize control\cite{NRF}.  The results of this
method were cross-checked against the results of using Richardson
extrapolation with Bulirsch-Stoer stepping\cite{NRF} and were found to
agree to the limits of the expected accuracy of either solution.}

	Experimental values for the low-energy Yukawa couplings are
extracted from the physical masses and mixing angles compiled by the
Particle Data Group\cite{PDG}, where entries of $Y_U$ are obtained by
dividing quark masses by $v \sin \beta /\sqrt{2}$ and those of
$Y_{D,L}$ by dividing quark and lepton masses by $v \cos \beta /\sqrt{2}$
, where $v = 246$ GeV.

	The experimental uncertainties on the observables (or
estimates for the quark masses) used in the fits are either those
appearing in Ref.~\cite{PDG} or $1\%$ of the central value, whichever
is larger; since the lepton masses are measured with extraordinary
precision, they are sensitive to the two-loop RGE and threshold
corrections that we have ignored.

The RGE for the neutrino Majorana mass matrix $M_{LL}$ was
computed in Ref.~\cite{BLP} and is included here in order to
complete the RGE evolution for all observables. The
low-energy neutrino observables are taken to be
\begin{eqnarray} \label{oldneut}
& & 100 < \frac{\Delta m_{23}^2}{\Delta m_{12}^2} < 2500, \nonumber \\
& & \sin^2 2\theta_{23} > 0.8, \nonumber \\
& & 2 \times 10^{-3} < \sin^2 2\theta_{12} < 10^{-2}.
\end{eqnarray}
For the sake of having meaningful uncertainties, a parameter whose
lower bound is much smaller than its upper bound is converted into its
logarithm.  Instead of Eq.~(\ref{oldneut}), we use
\begin{eqnarray}
\ln \left( \frac{\Delta m_{23}^2}{\Delta m_{12}^2} \right) & = & 6.22
\pm 1.61 , \nonumber \\
\sin^2 2\theta_{23} & = & 0.9 \pm 0.1 , \nonumber \\
\ln \left( \sin^2 2\theta_{12} \right) & = & -5.41 \pm 0.80 .
\end{eqnarray}
	Summarizing to this point, we have discussed the details of
how inputs consisting of the gauge couplings at $M_Z$ and Yukawa
matrix parameters at a high scale are manipulated using one-loop
RGE's to produce output values for fermion masses and mixing angles
observed at low energy.  Of course, the salient question is whether
one can find a choice of parameters $k_i$, where all of these
coefficients are $O(1)$, and yet the output quantities are all in
agreement with their observed values.\footnote{We also allow for
variation of the parameters $\epsilon$, $\epsilon^\prime$, $\rho$, and
$\xi$ by hand, but do not minimize with respect to them.
Changes in these parameters are equivalent to redefinitions of the
$O(1)$ coefficients, so that they merely set the scale for the
other parameters of the fit.}  This is accomplished through a $\chi^2$
minimization; thus, the complete simulation consists of choosing a set
of parameters $k_i$ (relevant at $M_{\rm GUT}$), running the RGE's down to
$m_t$, and comparing with observation to compute a figure of merit,
$\chi^2$.  If $\chi^2$ is too large, the parameters $k_i$ are adjusted
and the procedure is repeated until convergence of $\chi^2$ to a
minimum is achieved.

	The $\chi^2$ function assumes a somewhat nonstandard form.
Fermion masses and mixing angles are converted to Yukawa couplings
$k_i^{\rm expt} \pm \Delta k_i$, and contribute an amount
\begin{equation}
\Delta \chi^2 = \left( \frac{k_i^{\rm expt} - k_i}{\Delta k_i}
\right)^2
\end{equation}
to $\chi^2$, as usual.  There are 15 observables (6 quark masses, 3
quark CKM elements [since CP violation is neglected], 3 lepton masses,
2 neutrino mixing angles, and 1 neutrino mass ratio) and 26 parameters
$k_i$; on the surface, it seems that the fit is always
under-constrained.  However, our demand that the parameters $k_i$ lie
near unity imposes additional restrictions, which we include by adding
terms to $\chi^2$ of the form
\begin{equation}
\Delta \chi^2 = \left( \frac{\ln |k_i|}{\ln 3} \right)^2
\end{equation}
for each $i$.  Thus, the parameters $k_i$ are effectively no longer
free, but are to be treated analogously to pieces of data, each of
which contributes one unit to $\chi^2$ if it is as large as 3 or as
small as 1/3.  The particular choice of 3 for this purpose is, of
course, a matter of taste.  In effect, the inclusion of such terms renders
the parameters $k_i$ no longer as true degrees of freedom.  On the
other hand, they are not true pieces of data either, since a value of
say, $k_i = 0.8$ is just as valid as a value of $-1.1$ for our purposes.
Thus, the value of $\chi^2_{\rm min}$ determining a `good' fit is
15, since there are 15 pieces of true data and effectively no
{\em unconstrained\/} fit parameters.

	The numerical minimization is carried out using the MINUIT
minimization package. As a cross check, minimization using
Powell's direction set method\cite{NRF} is carried out to make sure
that the same minimum is achieved.  Since the topography of the
$\chi^2$ function is complicated due to the numerous parameters
involved, it is important to try a number of initial choices for the
input parameters $k_i$ in order to have confidence that the minimum
obtained is close to global.  Once convergence is achieved, a
parabolic minimum is assumed and a Hessian matrix is computed in order
to gauge uncertainties of the parameters.

	Detailed numerical fits show that it is not difficult to find
parameters $k_i$ that satisfy the constraint $\chi^2_{\rm min} < 15$.
However, in the $T^\prime \times Z_3$ model, the ratio $m_b/m_t$ must
be accommodated either by a small value of $\xi$ or a large value of
$\tan \beta$.  For definiteness, we choose $\tan \beta = 2$
as a representative value, and find a best fit with $\chi^2_{\rm min}$
of 2.77.  The complete set of parameters is listed in
Table~\ref{fit} and a comparison to data appears in Table~\ref{data}.  
Note especially that the parameters $\epsilon$,
$\epsilon^\prime$, and $\rho$ are somewhat larger (a factor of 2 or
more) than their values in the U(2) model of Ref.~\cite{u22}, 
where neutrino physics was not considered.  From
the excellent $\chi^2$, one concludes that the $T^\prime \times Z_3$
model has little difficulty satisfying all of the required constraints
including the naturalness of the coefficients, allowing for the small
parameter $\xi$ that distinguishes the scale of $Y_U$ from $Y_{D,L}$.

	While we have seen that the minimal scenario is extremely
 successful at reproducing 
fermion masses and mixing angles, there are nonetheless a number of
interesting variant models based on $T'$ symmetry. We
explore these models in the next three sections.

\section{SU(5)$\times$U(2) with neither SU(5) nor U(2)} \label{su5u2}
As discussed in the Introduction, the U(2) model must be embedded in a grand
unified theory to reproduce all of the observed quark mass hierarchies. 
In this section we present a model that does exactly the same, without the
need for a GUT, by extending the discrete gauged flavor group to
$T' \times Z_6$. We show that this model explains the ratio
$m_b/m_t$, which is merely parameterized in the U(2) model (and in
our other $T'$ models).
Before presenting the model we comment on notation. As before,
we use the triality superscripts $+, -$, and $0$ for the different 
representations of $T'$. For the $Z_{6}$ reps
we now introduce the indices $i = 0, 1, \dots ,5$, which 
combine through addition modulo 6. 
For example, ${\bf 2}^{+4} \otimes {\bf 1}^{+2} = {\bf 2}^{-0}$, 
etc. Since $Z_6$ is isomorphic to $Z_3 \times Z_2$, one may view the
new flavor symmetry as a $Z_2$ extension of the $T' \times Z_3$
flavor group defined in the model of Section~\ref{minimal};
denoting the $Z_2$ reps by $+$ and $-$, one identifies
\begin{eqnarray}
\nonumber
\begin{tabular}{||c|c||c||}
\hline
\hline
\,\, $Z_3$ \,\, & \,\, $Z_2$ \,\, & \,\, $Z_6$ \,\,\\
\hline
$0$ & $+$ & $0$ \\
$-$ & $-$ & $1$ \\
$+$ & $+$ & $2$ \\
$0$ & $-$ & $3$ \\
$-$ & $+$ & $4$ \\
$+$ & $-$ & $5$ \\
\hline
\hline
\end{tabular}
\end{eqnarray}
That is, the $Z_6$ charge is $2 \times (Z_3 \,\,{\rm charge}) + 3 \times
(Z_2 \,\,{\rm charge}) \,\, {\rm modulo} \,\, 6$.
In the remainder of this section we use the more compact 
$T' \times Z_{6}$ notation. 

The three generations of matter fields transform as
\begin{eqnarray} \label{QUD}
Q,U,D \sim {\bf 2}^{04} \oplus {\bf 1}^{00} \,\, ,
\end{eqnarray}
\begin{eqnarray} \label{L}
L \sim {\bf 2}^{04} \oplus {\bf 1}^{+4} \,\, ,
\end{eqnarray}
\begin{eqnarray} \label{E}
E \sim {\bf 2}^{+2} \oplus {\bf 1}^{-2} \,\, ,
\end{eqnarray}
\begin{eqnarray} \label{N}
\nu_{R} \sim {\bf 2}^{04} \oplus {\bf 1}^{+1} \,\, .
\end{eqnarray}
The matter fields have transformation properties that differ from those
in our previous models, and in
particular, the electroweak doublet leptons are no longer anomaly
free by themselves. The third-generation $L$ field is assigned to a
nontrivial $T'$ singlet, the ${\bf 1}^{+}$, which does not form a
complete SU(2) representation. Given the discussion in Section~\ref{gauge},
the $T'$ SU(2)$_{W}^2$ anomaly is not automatically cancelled. However,
we remedy this problem by assigning non-trivial transformation properties
to the Higgs fields:
\begin{eqnarray} \label{higgs}
H_{U} \sim {\bf 1}^{00} , \,\,  H_{D} \sim {\bf 1}^{-2}.
\end{eqnarray}
The fields $H_D$ and $L_3$ are both electroweak doublets and, as far
as the non-Abelian quantum numbers are concerned, form a vector-like pair
when $H_D$ is a ${\bf 1}^{-}$ under $T'$. The remaining fields, $E$ and 
$\nu_{R}$, do not transform under any unbroken non-Abelian continuous
gauge groups and thus their $T' \times Z_6$ quantum numbers may be
assigned freely. 

In order to break the flavor symmetry and obtain the
fermion mass matrices we introduce the following flavons:
\begin{eqnarray} \label{flavons1}
S \sim {\bf 3}^{0}, \,\, A \sim {\bf 1}^{-0}, \,\, \phi \sim {\bf 2}^{02},
\end{eqnarray}
\begin{eqnarray} \label{flavons2}
\Delta \sim {\bf 1}^{+4}, \,\, \Delta^{'} \sim {\bf 1}^{-2} \,\, .
\end{eqnarray}
In addition to these flavon fields, we introduce two more in the neutrino 
sector of the theory. Their transformation properties are such that 
they do not alter the form of the
charged fermion Yukawa textures:
\begin{eqnarray} \label{nflvons}
\phi_{\nu} \sim {\bf 2}^{+3}, \,\, \Delta_{\nu} \sim {\bf 1}^{+1}\,\,\,.
\end{eqnarray}
Together with $\nu_{R}$, these fields 
are the only ones that transform nontrivially under the $Z_{2}$ 
subgroup of $Z_{6}$ (i.e., the only ones with odd $Z_6$ charges).
Again,
we are interested in a two-step breaking:
\begin{eqnarray} \label{breaking}
T' \times Z_{6} \stackrel{\epsilon}{\longrightarrow} Z_{3}^{D}\stackrel
{\epsilon'} {\longrightarrow} nothing,
\end{eqnarray}
where $Z_{3}^{D}$ is precisely the same subgroup as in the minimal $T' \times
Z_3$ model. Thus, by the same arguments presented in Section~\ref{minimal}, we
obtain the following patterns of vevs:

\begin{eqnarray} \label{vevsz6}
\frac{\langle S \rangle}{M_f} \sim  \left( \begin{array}{cc}
0 & 0 \\ 0 & \epsilon
\end{array} \right), \,\, \frac{\langle A \rangle}{M_f} \sim 
\left( \begin{array}{cc} 0 &
\epsilon' \\ -\epsilon' &0 \end{array} \right), \\
\frac{\langle \phi \rangle}{M_f} \sim \sigma_{2} \left( 
\begin{array}{c} 0 \\ \epsilon \end{array} \right), \,\,
\frac{\langle \Delta \rangle}{M_f} \sim \epsilon, \,\, 
\frac{\langle \Delta^{'} \rangle}{M_f} \sim \epsilon, \\
\frac{\langle \phi_{\nu} \rangle}{M_f} \sim \sigma_2 \left( 
\begin{array}{c} \epsilon' \\ \epsilon \end{array} 
\right), \,\, \frac{\langle \Delta_{\nu} \rangle}{M_f} \sim \epsilon.
\end{eqnarray} 
Unlike the minimal model described in the previous two sections,
the flavons here contribute to the Yukawa matrices in some cases
only at quadratic order
\begin{eqnarray} \label{yuz6}
Y_{U} & \sim & \left( \begin{array}{c|c}
[ {\bf 3}^{4}\oplus {\bf 1}^{04}]
& [ {\bf 2}^{02}] \\
\hline
{[} {\bf 2}^{02}]
 & [ {\bf 1}^{00} ]
\end{array} \right)  \sim 
\left( \begin{array}{c|c}
\Delta S + \Delta A + \phi^2
& \phi \\
\hline
\phi
 & 1
\end{array} \right) 
\approx\left( \begin{array}{ccc}
0 & \epsilon \epsilon' & 0 \\
-\epsilon \epsilon' & \epsilon^{2} & \epsilon \\
0 & \epsilon & 1
\end{array} \right) \,\, ,
\end{eqnarray}
\begin{eqnarray} \label{ydz6}
Y_{D} & \sim & \left( \begin{array}{c|c}
[ {\bf 3}^{2}\oplus {\bf 1}^{+2}]
& [ {\bf 2}^{+0}] \\
\hline
{[} {\bf 2}^{+0}]
 & [ {\bf 1}^{+4} ]
\end{array} \right)  \sim 
\left( \begin{array}{c|c}
\Delta^{'} S + \Delta^{'} A
& \Delta \phi \\
\hline
\Delta \phi
 & \Delta
\end{array} \right)
\approx\left( \begin{array}{ccc}
0 & \epsilon' & 0 \\
-\epsilon' & \epsilon & \epsilon \\
0 & \epsilon & 1
\end{array} \right)\epsilon \,\, ,
\end{eqnarray}
\begin{eqnarray} \label{ylz6}
Y_{L} & \sim & \left( \begin{array}{c|c}
[ {\bf 3}^{4}\oplus {\bf 1}^{04}]
& [ {\bf 2}^{-4}] \\
\hline
{[} {\bf 2}^{-4}]
 & [ {\bf 1}^{+4} ]
\end{array} \right)  \sim 
\left( \begin{array}{c|c}
\Delta S + \Delta A + \phi^2
& \Delta^{'} \phi + \Delta_{\nu} \phi_{\nu}\\
\hline
\Delta^{'} \phi + \Delta_{\nu} \phi_{\nu}
 & \Delta
\end{array} \right)
\approx\left( \begin{array}{ccc}
0 & \epsilon' & \epsilon' \\
-\epsilon' & \epsilon & \epsilon \\
\epsilon' & \epsilon & 1
\end{array} \right)\epsilon \,\, .
\end{eqnarray}

We see that the flavons $\Delta$ and $\Delta^{'}$ appear in precisely
the right way to recover approximate SU(5) $\times$ U(2) textures for
$Y_D$ and $Y_L$, {\em with an additional overall factor of $\epsilon$}.
The only difference is a relatively uninteresting $\epsilon'$ entry
in the 13 and 31 elements of $Y_L$. Notice that the vev 
of the $\Sigma$ field has been replaced by 
$\langle \Delta \rangle$ in Eq.~(\ref{yuz6}). Thus, all important
features of the SU(5) $\times$ U(2) model are reproduced.

	Note that the ratio $m_{b}/m_{t}$,
which is experimentally observed to be in the range $ 0.023 \alt 
m_{b}/m_{t} \alt 
0.026 $, is predicted to be of order $\epsilon \approx 0.02$ for
$\tan \beta \approx O(1)$, as can be seen from the 
ratio of the 33 entries in $Y_U$ and $Y_D$. This is promising since
$\tan \beta \approx O(1)$ is the naive expectation if the Higgs potential is not
fine-tuned.

	Before proceeding to the analysis of the neutrino sector, a
few comments are warranted on the possible supersymmetric contributions
to FCNC's in this model. As mentioned in the Introduction, scalar
superpartners of the first two generations are exactly degenerate
in our models when the flavor symmetry is unbroken. The amount of
scalar nondegeneracy at low energies is determined by the order at which
flavons contribute to the scalar mass matrices. In the minimal model, the
flavons contribute quadratically to the scalar masses of the first two
generations, as a consequence of the flavons' nontrivial $Z_3$ charges.
The scalar mass-squared matrices of the U(2) model are then reproduced.
In the current model, however, the flavon $S$ may contribute 
linearly, since ${\bf 3}^{0}$ is in the product of $({\bf 2}^{04})^{\dagger} 
\otimes ({\bf 2}^{04})$. The important point is that this effect provides
an $O(\epsilon)$ correction to the {\em diagonal} entries of the scalar
mass matrices. In the fermion mass-eigenstate basis, a Cabibbo-like
rotation $\theta_{C} \sim \epsilon'/\epsilon$ leads to 12 entries in
the scalar mass matrices of order $\epsilon' \tilde{m}_{0}^{2}$, where
$\tilde{m}_{0}^{2}$ is an average scalar mass, and $\epsilon' \approx
0.004$. Taking into account uncertainty in $O(1)$ coefficients, this
result is in marginal agreement with the bounds from CP-conserving
flavor-changing processes, assuming superpartner masses less than a
TeV~\cite{mas}. While bounds from CP-violating precesses are generically
stronger, the $O(1)$ coefficients have unknown phases that one may
simply choose in order to avoid these bounds. Without a firm
understanding  of the origin of CP violation, saying more about these
phases entails a degree of speculation that we choose to avoid.
Of course, if scalar superpartners are heavy (as in the `more minimal
MSSM'~\cite{nelson}) or flavor universal (as in gauge mediation~\cite{DN},
anomaly mediation~\cite{randall,luty}, 
or Scherk-Swartz mechanism~\cite{dim}) the
current $T'$ model is completely safe.

	Next we examine the neutrino sector of
the model. Given the transformation properties of $\nu_{R}$, 
we calculate the neutrino
Dirac and Majorana mass matrices
\begin{eqnarray} \label{mlrz6}
M_{LR} \sim \left( \begin{array}{c|c} [ {\bf 3}^{4} \oplus {\bf 1}^{04}]
& [ {\bf 2}^{-1}] \\
\hline
{[} {\bf 2}^{-4}]
 & [ {\bf 1}^{+1} ]
\end{array} \right) \sim 
\left( \begin{array}{c|c} \Delta S + \Delta A + \phi^2 & \Delta \phi_{\nu} \\
\hline
\Delta^{'} \phi + \Delta_{\nu} \phi_{\nu} & \Delta_{\nu}
\end{array} \right) \langle H_U \rangle \nonumber \\
\approx \left( \begin{array}{ccc} 0 & l_1 \epsilon' & l_2 r_1 \epsilon' \\
-l_1 \epsilon' & l_3 \epsilon & l_2 r_3 \epsilon \\
l_4 \epsilon' & l_5 \epsilon & l_6
\end{array} \right) \epsilon \langle H_U \rangle \,\, ,
\end{eqnarray}

\begin{eqnarray} \label{mrrz6}
M_{RR} \sim \left( \begin{array}{c|c} [ {\bf 3}^{4} ]
& [ {\bf 2}^{-1}] \\
\hline
{[} {\bf 2}^{-1}]
 & [ {\bf 1}^{+4} ]
\end{array} \right) \sim \left( \begin{array}{c|c} \Delta S  & 
\Delta \phi_{\nu} \\
\hline
\Delta \phi_{\nu} & \Delta
\end{array} \right) \Lambda_R
 \approx \left( \begin{array}{ccc} 0 & 0 &  r_1 \epsilon' \\
0 & r_2 \epsilon &  r_3 \epsilon \\
r_1 \epsilon' & r_3 \epsilon & r_4
\end{array} \right) \epsilon \Lambda_R \,\, ,
\end{eqnarray}
where $r_i$ and $l_i$ are $O(1)$ coefficients. To leading order, the seesaw
mechanism gives
\begin{eqnarray} \label{mllz6}
M_{LL} \sim \left( \begin{array}{ccc} {\epsilon'}^{2}/\epsilon & 
\epsilon' & \epsilon' \\
\epsilon' & 1 & 1 \\
\epsilon' & 1 & 1
\end{array} \right) \frac{\epsilon \langle H_U \rangle^{2}}{\Lambda_R} \,\, .
\end{eqnarray}
Note that the texture in Eq.~(\ref{mllz6}) is not changed if higher-order
corrections are included that lift the zeroes in 
Eqs.~(\ref{mlrz6})$-$(\ref{mrrz6}).
Following the same procedure as before, we diagonalize $M_{LL}$ and $Y_L$
and extract the neutrino masses and mixing angles. 
A global fit of the parameters in this model can in principle be done; 
however we just present a viable set of parameters for simplicity. 
Using the set of values for the $O(1)$ coefficients in $M_{LL}$
( $r_1, \dots, r_4, l_1, \dots,l_6) =$ (1.0, 1.0, 1.0, $-$1.0, 1.2,
1.2, 1.3, $-$1.0, $-$2.0, 1.0) and assuming all coefficients in $Y_L$ are 1.0 
except that of the 22 entry, which we set to 3.0, we obtain
\begin{equation}
\frac{\Delta m_{23}^{2}}{\Delta m_{12}^{2}} = 125 ,
\,\,\,\,\,\,\,\,\,\,
{\sin^{2} 2 \theta_{12}} = 3.5 \times 10^{-3} ,
\,\,\,\,\,\,\,\,\,\,
{\sin^{2} 2 \theta_{23}} = 0.88.
\end{equation}
This agrees with the allowed ranges described in the previous sections.
It is worth mentioning that the texture Eq.~(\ref{mllz6}) is the same as
obtained in Ref.~\cite{caronehall}, and thus the claim in Ref.~\cite{hallweiner} 
that this texture cannot account for solar neutrino oscillations is
not correct.
\section{A Global $T^\prime$ Model} \label{global}

As pointed out in the Introduction, it is not possible to construct a 
realistic supersymmetric model with a continuous SU(2) flavor symmetry
if scalar universality is not assumed.
The argument is straightforward: The left- and right-handed up quark 
fields must be embedded in {\bf 2}$\oplus${\bf 1} representations to 
maintain the heaviness of the top quark, as well as degeneracy of
squarks of the first two generations. Given this assignment,
the coupling $Q^a U^b \epsilon_{ab} H_u$ is allowed by the unbroken
flavor symmetry, which implies the unacceptable relation 
$m_u = m_c \approx m_t$.  The $T'$ model below demonstrates that
discrete subgroups of SU(2) are viable for building models of fermion 
masses, although they are more dangerous than models with additional 
Abelian factors, as far as supersymmetric FCNC processes 
are concerned.  We first present the model, and then explain how it 
evades the problem described above.

The crucial feature that allows one to build a successful $T'\times Z_3$
model is the existence of a doublet representation ${\bf 2}^{0-}$, whose 
first generation component alone rotates by a phase under the  $Z_3^D$ subgroup.  
This choice is unique in models where $T'$ is a discrete gauge
symmetry, since the ${\bf 2}^0$ rep is the only doublet that fills a 
complete SU(2) representation if we embed $T'$ in SU(2).  The
{\bf 4} of SU(2) decomposes into the reps ${\bf 2}^+$ and ${\bf 2}^-$,
which implies that each is separately anomalous.  While 
it might still be possible to construct models involving anomaly-free
combinations of ${\bf 2}^+$ and ${\bf 2}^-$ reps, we have found no examples
that are particularly compelling.  On the other hand, if $T'$ is assumed 
to be a global symmetry, then matter fields can be assigned to any of
the doublet representations freely.  This provides an opportunity 
for constructing economical models, as we now demonstrate.

Consider the $Z_3$ subgroup of $T'$ generated by the element $g_9$ that 
acts on the ${\bf 2}^{0}$ rep as $diag\{\eta^2,\eta\}$, with 
$\eta$ defined as in Section~\ref{tprime}.  In the ${\bf 2}^-$ rep, this
element takes the form $diag\{\eta, 1\}$, which we identify
as the desired phase rotation matrix for matter fields of the first
two generations.  Given our freedom to assign matter fields to any 
of the doublet reps in a global $T'$ model, it is no longer necessary
to extend the flavor symmetry by an Abelian factor in order to find 
a subgroup that forbids the order $\epsilon'$ Yukawa entries.  Thus, one is
naturally led to the charge assignment
\begin{equation}
\psi \sim {\bf 2}^- \oplus {\bf 1}^0 \,\,\, \mbox{ for } \,\,\,
\psi=Q, U, D, L \mbox{ and } E, 
\end{equation}
and $H_{U,D} \sim {\bf 1}^0$, which yields
\begin{eqnarray} \label{Yuk}
Y_{U,D,L} & \sim & \left( \begin{array}{c|c} 
[ {\bf 3}\oplus {\bf 1}^{-}] 
& [ {\bf 2}^{+}] \\ 
\hline
{[} {\bf 2}^{+}]
 & [ {\bf 1}^{0} ]
\end{array} \right) .
\label{eq:ytp}
\end{eqnarray}
Introducing flavons, $A$, $\phi$ and $S$ transforming as
${\bf 1}^{-}$, ${\bf 2}^{+}$, and ${\bf 3}$, respectively, one
reproduces the canonical U(2) textures assuming the
breaking pattern
\begin{eqnarray} \label{eq:nbreak}
T'  \stackrel{\epsilon}{\longrightarrow} Z_{3} 
\stackrel{\epsilon'}
{\longrightarrow} nothing,
\end{eqnarray}
together with the dynamical assumption that only the ${\bf 1}^-$ rep 
participates in the last step of symmetry breaking.  The resulting \
textures are identical to those in our original model of Section~\ref{minimal}.  
One difference, however, is that the $S$ flavon in this model 
contributes to the squark mass matrices at first order in $\epsilon$,
just as in the $T' \times Z_6$ model. However, this is not a concern
for the same reasons discussed at length in Section~\ref{su5u2}.

Turning to neutrino physics, recall that successful results 
were obtained in the $T'\times Z_3$ model by altering the charge 
assignment of the third-generation right-handed neutrino field.  
Thus, we are motivated here to consider 
\begin{equation}
\nu_R \sim {\bf 2}^- \oplus {\bf 1}^- \,\,\, ,
\end{equation}
which implies 
\begin{equation}
M_{LR} \sim  \left( \begin{array}{c|c} [ {\bf 3} \oplus {\bf 1}^{-}] 
& [ {\bf 2}^{-}] \\ 
\hline
{[} {\bf 2}^{+}]
 & [ {\bf 1}^{+} ]
\end{array} \right) \,\,\, , \hspace{2em} 
M_{RR} \sim \left( \begin{array}{c|c} [ {\bf 3} ] 
& [ {\bf 2}^{-}] \\ 
\hline
{[} {\bf 2}^{-}]
 & [ {\bf 1}^{-} ]
\end{array} \right) \,\,\, .
\end{equation}
We identify the flavon $\phi_\nu$ with the representation 
${\bf 2}^-$, which does not appear in any of the charged
fermion Yukawa textures.  However, there is an important
difference between this model and the one discussed in 
Section~\ref{minimal}: The third generation $\nu_R$ field transforms by a
phase under the $Z_3$ subgroup, so that, for example,
the 13 and 31 entries of $M_{RR}$ are left invariant under 
this intermediate symmetry.  This implies an 
inversion in the hierarchy of vevs in the third row and column of 
$M_{RR}$.  In the non-unified version of the model, it is somewhat
remarkable that we still obtain a viable form for $M_{LL}$:
\begin{equation}
M_{LR}  \approx  \left( \begin{array}{ccc} 0 
& l_1\epsilon' & l_5 r_1\epsilon \\ 
-l_1\epsilon' & l_2 \epsilon & l_3 r_2 \epsilon' \\ 0 &
l_4 \epsilon & 0 \end{array} \right) \langle H_U \rangle \,\,\, ,
\hspace{2em}
M_{RR}  \approx  \left( \begin{array}{ccc}  \ 0
& 0 & r_1 \epsilon \\ 
0 & r_3 \epsilon & r_2 \epsilon' \\ r_1 \epsilon &
r_2 \epsilon' & r_4 \epsilon' \end{array} \right) \Lambda_R \,\,\, ,
\end{equation}
\begin{eqnarray} 
M_{LL}  \sim  \left( \begin{array}{ccc} (\epsilon'/\epsilon)^2 & 
\epsilon'/\epsilon & \epsilon'/\epsilon 
\\ \epsilon'/\epsilon & 1 & 1 
\\ \epsilon'/\epsilon & 1 & 1
\end{array} \right) \frac{\langle H_U \rangle^2 \epsilon}{\Lambda_R} ,
\end{eqnarray}
Unfortunately, this result does not persist in the simplest
unified version of the model, which includes additional suppression
factors in the 22 entries of $M_{LR}$ and $M_{RR}$.  Fortunately, a simple 
modification of the flavon charge assignments in the unified theory 
allows us to recover the previous result.   We introduce two $\phi_\nu$ flavons 
that transform differently under $T' \times$ SU(5):
\begin{equation}
\phi_\nu \sim ({\bf 2}^-, {\bf 24})  \,\,\,\,\, \mbox{,}
\,\,\,\,\, \phi_\nu' \sim  ({\bf 2}^-, {\bf 1})  \,\,\, .
\end{equation}
Furthermore, we assume the pattern of vevs
\begin{equation}
\langle \phi_\nu\rangle \sim \left(\begin{array}{c} 0 \\ \epsilon
\end{array}\right) \,\,\, , \,\,\,
\langle \phi'_\nu\rangle \sim \left(\begin{array}{c} \epsilon' \\ 0
\end{array}\right) \,\,\, .
\end{equation}
This is consistent with the breaking pattern in Eq.~(\ref{eq:nbreak}), but 
includes a dynamical assumption that the doublet $\phi'_\nu$ does not
participate in the first stage of sequential symmetry breaking and its
second component acquires no vev.\footnote{We consistently assume that a
flavon that transforms nontrivially under a subgroup $H_i$ either acquires
a vev of order the scale at which $H_i$ is spontaneously broken, or
acquires no vev at all.}  
Since $\phi_\nu$ transforms as an SU(5) adjoint, it can contribute directly
to $M_{LR}$, but only to $M_{RR}$ if, for example, the adjoint flavon $\Sigma$ is 
also present; the corresponding entries of $M_{RR}$ are therefore suppressed by an 
additional factor of $\epsilon$:  
\begin{equation}\label{eq:gtex}
M_{LR}  \approx  \left( \begin{array}{ccc} 0 
& l_1\epsilon' & l_5 r_1\epsilon \\ 
-l_1\epsilon' & l_2 \epsilon^2 & l_3 r_2 \epsilon' \\ 0 &
l_4 \epsilon & 0 \end{array} \right) \langle H_U \rangle \,\,\, ,
\hspace{2em}
M_{RR}  \approx  \left( \begin{array}{ccc}  \ 0
& 0 & r_1 \epsilon^2 \\ 
0 & r_3 \epsilon^2 & r_2 \epsilon' \\ r_1 \epsilon^2 &
r_2 \epsilon' & r_4 \epsilon' \end{array} \right) \Lambda_R \,\,\, .
\end{equation}
The seesaw mechanism then yields
\begin{eqnarray} \label{eq:gmll}
M_{LL}  \sim  \left( \begin{array}{ccc} (\epsilon'/\epsilon)^2 & 
\epsilon'/\epsilon & \epsilon'/\epsilon 
\\ \epsilon'/\epsilon & 1 & 1 
\\ \epsilon'/\epsilon & 1 & 1
\end{array} \right) \frac{\langle H_U \rangle^2 \epsilon}{\Lambda_R} ,
\end{eqnarray}
where we used the numerical fact that $\epsilon'^2/\epsilon^3 \sim O(1)$.
It is important to note that we have only displayed the contributions to 
Eq.~(\ref{eq:gtex}) linear in $\phi$, $S$ and $A$, for convenience; 
quadratic and higher order corrections lift the zero entries of 
these textures, but do not change the result in Eq.~(\ref{eq:gmll}) 
qualitatively.  Note that Eq.~(\ref{eq:gmll}) is the same successful 
texture obtained in our original $T'\times Z_3$ model. 

Finally, we return to the no-go theorem presented at the beginning
of this section.  It is not possible to construct a realistic
model with a continuous SU(2) flavor symmetry and
${\bf 2} \oplus {\bf 1}$ rep structure  because an unwanted
flavor-invariant operator may be formed from the product of
two doublet matter fields.  In our global $T'$ model we have
the freedom to assign matter fields to new doublet representations 
whose products contain no trivial singlets, thus avoiding the problem.

\section{$T'$ with Sterile Neutrinos} \label{sterile}

In this section we comment briefly on the possibility of four light 
neutrino species.  Rather than investigating the (vast) space of possible models,
we simply show how the results of a successful extension of the U(2) model with 
a sterile neutrino proposed by Hall and Weiner (HW)~\cite{hallweiner}
can be reproduced with $T'$ symmetry instead.

Consider a U(2) model with all matter fields, including three generations 
of right-handed neutrinos, in ${\bf 2}\oplus{\bf 1}$ representations.  
Given the canonical pattern of flavon vevs, one  obtains a right-handed 
neutrino mass matrix of the form
\begin{equation}\label{eq:mrrsing}
M_{RR}  =  \left( \begin{array}{ccc}  
0  &  0  &  0  \\ 
0  & \epsilon & \epsilon \\ 
0  & \epsilon & 1  \end{array} \right) \Lambda_R \,\,\, .
\end{equation}
Since $M_{RR}$ is symmetric, there is no contribution from the
flavon $A$, leading to a singular matrix.  It is important to
emphasize that the zero entries of Eq.~(\ref{eq:mrrsing}) are
not lifted at any order in $\epsilon$ and $\epsilon'$ as a consequence 
of the holomorphicity of the superpotential.  From consideration of 
the U(2) indices of the flavon fields (or alternatively their charges 
under a U(1) subgroup of U(2)), it is possible to show that any 
contribution to the vanishing entries of Eq.~(\ref{eq:mrrsing}) requires 
the complex conjugation of a flavon field, which is not allowed by unbroken 
supersymmetry.  If the pattern of flavon vevs is not altered, the 
first-generation right-handed neutrino remains in 
the low-energy theory as a sterile neutrino.

This sterile neutrino mixes with the second-generation left-handed
neutrino at order $\epsilon'$ in $M_{LR}$.  After integrating out
the two heavy right-handed neutrino flavors, one obtains a four-by-four
neutrino mass matrix of the form
\begin{equation}
M^{(4)} = \left( \begin{array}{ccc|c} &&& 0\\
& M^{(3)}_{LL} & & c \epsilon' \langle H_U \rangle\\	
&&& 0 \\\hline
0 & c \epsilon' \langle H_U \rangle & 0 &0  \end{array} \right)  \,\,\, ,
\end{equation}
where the three-by-three block $M^{(3)}_{LL}$ has entries of order
$\langle H_U \rangle^2/\Lambda_R$, which can be found in 
Ref.~\cite{hallweiner}.  HW observe that the 24 and 42 entries of $M^{(4)}$ 
are much larger than all others, leading naturally to maximal mixing 
between $\nu_\mu$ and the sterile neutrino.  As it stands, however, both would 
have masses of order of the electroweak scale unless $c$ is taken to be
of $O(10^{-8})$.  To obtain a viable model, HW extend the flavor
symmetry by an additional U(1) factor, under which all the right-handed
neutrinos have charge $+1$. A charge $-1$ flavon is introduced 
with the vev $\epsilon_N\sim 10^{-8}$, which breaks this symmetry weakly.
One then finds that $c\approx \epsilon_N$, while $M^{(3)}$ remains unchanged.  

The main obstacle to implementing this solution in a $T'\times Z_3$
model with all matter fields assigned to ${\bf 2}^{0-}\oplus {\bf 1}^{00}$
reps is that higher-order corrections to the first row and column of
Eq.~(\ref{eq:mrrsing}) are not forbidden by holomorphicity; the
complex conjugate of any non-trivial $Z_3$ phase rotation is the
same as its square.  Thus, we are led to promote our $Z_3$ 
symmetry to a continuous U(1).\footnote{We
could also promote $Z_3$ to a much larger $Z_n$ that adequately
suppresses corrections to the zero entries in Eq.~(\ref{eq:mrrsing});
we leave this possibility implicit in our discussion.}  The 
appropriate embedding is given by
\[
\psi \sim {\bf 2}^{0-} \oplus {\bf 1}^{00}
\longrightarrow  {\bf 2}^0_{+1} \oplus {\bf 1}^0_0
\]\begin{equation}
\phi \sim {\bf 2}^{0+} \longrightarrow {\bf 2}^0_{-1}
\,\,\,\,\, , \,\,\,\,\, S \sim {\bf 3}^- \longrightarrow {\bf 3}_{-2}
\,\,\,\,\, , \,\,\,\,\,A \sim {\bf 1}^{0-} \longrightarrow {\bf 1}^0_{-2} \, ,
\end{equation}
where the subscript indicates the U(1) charge.  Assuming the breaking
pattern
\begin{equation}
T'\times U(1) \stackrel{\epsilon}{\longrightarrow} Z^D_3 
\stackrel{\epsilon'}{\longrightarrow}  nothing,
\end{equation}
we reproduce the textures of the U(2) model, including Eq.~(\ref{eq:mrrsing}), 
identically.  The HW predictions for solar, atmospheric and LSND~\cite{lsnd}
neutrino oscillations are  then recovered by extending the symmetry by 
an additional U(1) factor, implemented precisely as before.  We are
thus able to reproduce the results of Ref.~\cite{hallweiner}
with the flavor symmetry $T'\times$U(1)$^2$.  Although we find this
model less compelling than the other three already
discussed, it may be of some relevance if the LSND oscillation
result is independently confirmed.

\section{Conclusions} \label{conclusions}

We have shown in this paper how to reproduce the quark and 
charged lepton Yukawa textures of the U(2) model using a {\em minimal} 
non-Abelian discrete symmetry, the double tetrahedral group
$T'$.  The first model we discuss, based on the discrete gauge 
symmetry $T'\times Z_3$, not only successfully accommodates the 
observed charged fermion masses and CKM angles, but also accounts 
for solar (small-angle MSW) and atmospheric neutrino oscillations.  
In particular, a large $\nu_\mu$-$\nu_\tau$  mixing angle is predicted 
in the model, even though all charged fermion Yukawa textures are 
hierarchical.  A global fit including neutrino parameters was 
performed in a grand unified version of the model, and results 
with extremely good $\chi^2$ were obtained. 

In addition, two variant $T'$ models were discussed.  In the first, the 
flavor group was extended to $T'\times Z_6$, and all important features
 of the 
SU(5)$\times$U(2) model were reproduced without the need for a 
field-theoretic unification.  This model provided a successful 
prediction (with order-one uncertainty) of the bottom to top 
Yukawa coupling ratio, which is merely parameterized in the U(2) model 
and in the other $T'$ models we discuss.  The second variant theory was 
based on a global $T'$ symmetry and demonstrates that the successful 
U(2) textures can be obtained without including an Abelian 
factor in the flavor group.  In both variant models, large  
$\nu_\mu$-$\nu_\tau$ mixing is predicted, and solutions to the solar 
and atmospheric neutrino problems are naturally obtained.

It is worth pointing out that the viable neutrino textures predicted
by our models are achieved without altering the predictive textures
of the charged fermions, and without introducing sterile neutrinos.  
Interestingly, the solutions we present have no 
simple analogy in the U(2) model: the right-handed neutrino
fields in our models do not fill complete U(2) representations.
In particular, the third generation $\nu_R$ transforms as a ${\bf 1}^-$,  
which forms only {\em part} of a ${\bf 5}$ in U(2).  Aside from the
possibility of very nonminimal U(2) models (e.g. with seven
generations of right-handed neutrinos), the desired neutrino $T'$ 
reps do not naturally occur.  The key advantage of discrete groups
is that the large, phenomenologically unused representations
of the continuous embedding group break up into sets of small
phenomenologically useful representations of the discrete group.
If discrete gauge symmetries arise as fundamental symmetries of nature, 
then we see from the example of $T'$ that their richer representation 
structure makes it possible to construct simple and elegant models of 
flavor.

{\samepage
\begin{center}
{\bf Acknowledgments}
\end{center}
AA and CDC thank the National Science Foundation for support under 
Grant Nos.\ PHY-9800741 and PHY-9900657, and the Jeffress Memorial
Trust for support under Grant No. J-532.  RFL thanks the Department of 
Energy for support under Contract No.\ DE-AC05-84ER40150.}
%\appendix

\appendix
\section{Explicit Details of $T^\prime$}

As described in the text, the group $T^\prime$ is generated by the
elements labeled $g_5$ and $g_9$.  We begin by exhibiting explicit
matrices representing these elements in each of the seven reps listed
in Table~I.  The singlets are $g_5({\bf 1}^{0,\pm}) = 1$, $g_9({\bf
  1}^0) = 1$, $g_9({\bf 1}^+) = \eta$, $g_9({\bf 1}^-) = \eta^2$,
where $\eta = \exp(2\pi i/3)$.  The doublets are
\begin{equation}
g_5 ({\bf 2}^{0,\pm}) = M_1 ,  \,\,\,\,
g_9 ({\bf 2}^0) = \eta M_2, \,\,\,\, 
g_9 ({\bf 2}^+) = \eta^2 M_2, \,\,\,\,
g_9 ({\bf 2}^-) = M_2 ,
\end{equation}
where
\begin{equation}
M_1 = -\frac{1}{\sqrt{3}} \left( \begin{array}{cc} +i & +\sqrt{2}
    e^{i\pi /12} \\ -\sqrt{2} e^{-i\pi /12} & -i \end{array} \right),
    \ \
M_2 = \left( \begin{array}{cc} \eta & 0 \\ 0 & 1 \end{array} \right) ,
\end{equation}
and the triplet rep is generated by
\begin{equation}
g_5 ({\bf 3}) = \frac 1 3 \left( \begin{array}{ccc} -1 & 2\eta &
    2\eta^2 \\ 2\eta^2 & -1 & 2\eta \\ 2\eta & 2\eta^2 & -1
    \end{array} \right) , \ \
g_9 ({\bf 3}) = \left( \begin{array}{ccc} 1 & 0 & 0 \\ 0 & \eta & 0
\\ 0 & 0 & \eta^2 \end{array} \right) .
\end{equation}
The Clebsch-Gordan (CG) coefficient matrices ${\cal O}_i$ coupling an
$n_x$-plet {\bf x} and an $n_y$-plet {\bf y} to form an $n_z$-plet
{\bf z} consist of $n_z$ matrices of dimensions $n_x \times n_y$
satisfying the condition
\begin{equation}\label{eq:a4}
R_x^T {\cal O}_i R_y = \sum_{j=1}^{n_z} (R_z)_{ij} {\cal O}_j, \ \ i =
1, \ldots , n_z ,
\end{equation}
where $R_i$ denotes the group rotation $R$ in rep $i$. In a perhaps more familiar 
notation, the CGs above may be written
\begin{equation} \label{clebsch}
\left( {\cal O}_i \right)_{jk} = \left( \begin{array}{cc} {\bf x} &
    {\bf y} \\ j & k \end{array} \right| \left. \begin{array}{c} {\bf
    z} \\ i \end{array} \right) .
\end{equation}
Note from Eq.~(\ref{clebsch}) that the CG matrices for 
${\bf R}_1 \otimes {\bf R}_2$ are simply the transposes of those 
for ${\bf R}_2 \otimes {\bf R}_1$, and thus are omitted below.  The 
coefficients $c$ below indicate multiplicative constants arbitrary in the definition
Eq.~(\ref{eq:a4}). The CG coefficients for two singlet reps
or any rep with ${\bf 1}^{0}$ are all
unity; the remaining CGs for products involving singlets are
\begin{equation}
{\bf 1}^{t_1} \otimes {\bf 2}^{t_2} = {\bf 2}^{t_1 + t_2} , \
{\rm with} \ 
{\cal O}_1 = c (1 \ 0), \ {\cal O}_2 = c (0 \ 1) .
\end{equation}
\begin{equation}
{\bf 1}^+ \otimes {\bf 3} = {\bf 3} , \ {\rm with} \
{\cal O}_1 = c (0 \ 0 \ 1), \ {\cal O}_2 = c (1 \ 0 \ 0), \ {\cal O}_3
= c (0 \ 1 \ 0) .
\end{equation}
\begin{equation}
{\bf 1}^- \otimes {\bf 3} = {\bf 3} , \ {\rm with} \
{\cal O}_1 = c (0 \ 1 \ 0), \ {\cal O}_2 = c (0 \ 0 \ 1), \ {\cal O}_3
= c (1 \ 0 \ 0) .
\end{equation}
Next, let 
\begin{eqnarray}
& &
M_3 = \frac 1 2 (1-i) \left( \begin{array}{cc} 0 & 1 \\ 1 & 0
\end{array} \right), \ \
M_4 = \left( \begin{array}{cc} i & 0 \\ 0 & 0 \end{array} \right),
\nonumber \\ & &
M_5 = \left( \begin{array}{cc} 0 & 0 \\ 0 & 1 \end{array} \right),
\ \
M_6 = \left( \begin{array}{cc} 0 & 1 \\ -1 & 0 \end{array}
\right).
\end{eqnarray}
Then
\begin{eqnarray}
\lefteqn{{\bf 2}^0 \otimes {\bf 2}^0 \supset {\bf 3}, \
{\bf 2}^\pm \otimes {\bf 2}^\mp \supset {\bf 3}:} & & \nonumber \\
& & {\cal O}_1 = c M_3 , \ {\cal O}_2 = c M_4, \ {\cal O}_3 = c M_5 .
\\ & & \nonumber \\
\lefteqn{{\bf 2}^0 \otimes {\bf 2}^0 \supset {\bf 1}^0, \
{\bf 2}^{\pm} \otimes {\bf 2}^{\mp} \supset {\bf 1}^0:} & &
\nonumber \\ & & {\cal O} = c M_6 .
\\ & & \nonumber \\
\lefteqn{{\bf 2}^0 \otimes {\bf 2}^+ \supset {\bf 3}, \
{\bf 2}^- \otimes {\bf 2}^- \supset {\bf 3}:} & & \nonumber \\
& & {\cal O}_1 = c M_5, \ {\cal O}_2 = c M_3, \ {\cal O}_3 = c M_4 .
\\ & & \nonumber \\
\lefteqn{{\bf 2}^0 \otimes {\bf 2}^+ \supset {\bf 1}^+, \
{\bf 2}^- \otimes {\bf 2}^- \supset {\bf 1}^+ :} & & \nonumber  \\
& & {\cal O} = c M_6 .
\\ & & \nonumber \\
\lefteqn{{\bf 2}^0 \otimes {\bf 2}^- \supset {\bf 3}, \ {\bf 2}^+
\otimes {\bf 2}^+ \supset {\bf 3}:} & & \nonumber \\ & &
{\cal O}_1 = c M_4, \ {\cal O}_2 = c M_5, \ {\cal O}_3 = c M_3 .
\\ & & \nonumber \\
\lefteqn{{\bf 2}^0 \otimes {\bf 2}^- \supset {\bf 1}^-, \ {\bf 2}^+
\otimes {\bf 2}^+ \supset {\bf 1}^-:} & &  \nonumber \\ & & 
{\cal O} = c M_6 .
\end{eqnarray}
The remaining combinations are:
\begin{eqnarray}
\lefteqn{ {\bf 2}^{0,\pm} \otimes {\bf 3} \supset
{\bf 2}^{0, \pm}:} & & \nonumber \\ & &
{\cal O}_1 = c \left( \begin{array}{ccc} 1 & 0 & 0 \\ 0 & 1+i & 0
  \end{array} \right), \ {\cal O}_2 = c \left( \begin{array}{ccc}
0 & 0 & 1-i \\ -1 & 0 & 0 \end{array} \right) .
\\ & & \nonumber \\
\lefteqn{{\bf 2}^0 \otimes {\bf 3} \supset {\bf 2}^+, \
{\bf 2}^+ \otimes {\bf 3} \supset {\bf 2}^-, \ {\bf 2}^- \otimes
{\bf 3} \supset {\bf 2}^0:} & & \nonumber \\ & &
{\cal O}_1 = c \left( \begin{array}{ccc} 0 & 1 & 0 \\ 0 & 0 & 1+i
  \end{array} \right), \ {\cal O}_2 = c \left( \begin{array}{ccc} 1-i
    & 0 & 0 \\ 0 & -1 & 0 \end{array} \right) .
\\ & & \nonumber \\
\lefteqn{{\bf 2}^0 \otimes {\bf 3} \supset {\bf 2}^-, \
{\bf 2}^+ \otimes {\bf 3} \supset {\bf 2}^0, \ {\bf 2}^- \otimes
{\bf 3} \supset {\bf 2}^+ :} & & \nonumber \\ & &
{\cal O}_1 = c \left( \begin{array}{ccc} 0 & 0 & 1 \\ 1+i & 0 & 0
  \end{array} \right) , \ {\cal O}_2 = c \left( \begin{array}{ccc} 0 &
    1-i & 0 \\ 0 & 0 & -1 \end{array} \right) .
\end{eqnarray}
\begin{eqnarray}
\lefteqn{{\bf 3} \otimes {\bf 3} \supset {\bf 3}_{s} \oplus {\bf 3}_{a}
:} & & \nonumber \\
{\cal O}_1 & = & c_1 \left( \begin{array}{ccc} 2 & 0 & 0 \\ 0 & 0 & -1
    \\ 0 & -1 & 0 \end{array} \right) + c_2 \left( \begin{array}{ccc}
0 & 0 & 0 \\ 0 & 0 & -1 \\ 0 & 1 & 0 \end{array} \right) , \nonumber \\
{\cal O}_2 & = & c_1 \left( \begin{array}{ccc} 0 & -1 & 0 \\ -1 & 0 &
    0 \\ 0 & 0 & 2 \end{array} \right) + c_2 \left( \begin{array}{ccc}
    0 & -1 & 0 \\ 1 & 0 & 0 \\ 0 & 0 & 0 \end{array} \right) , \nonumber \\
{\cal O}_3 & = & c_1 \left( \begin{array}{ccc} 0 & 0 & -1 \\ 0 & 2 & 0
    \\ -1 & 0 & 0 \end{array} \right) + c_2 \left( \begin{array}{ccc}
    0 & 0 & 1 \\ 0 & 0 & 0 \\ -1 & 0 & 0 \end{array} \right) .
\end{eqnarray}
\begin{eqnarray}
{\bf 3} \otimes {\bf 3} \supset {\bf 1}^0 : \ \ & & {\cal O} =
c \left( \begin{array}{ccc} 1 & 0 & 0 \\ 0 & 0 & 1 \\ 0 & 1 & 0
\end{array}
\right) , \nonumber \\
{\bf 3} \otimes {\bf 3} \supset {\bf 1}^+ : \ \ & & {\cal O} =
c \left( \begin{array}{ccc} 0 & 1 & 0 \\ 1 & 0 & 0 \\ 0 & 0 & 1
\end{array} 
\right) , \nonumber \\
{\bf 3} \otimes {\bf 3} \supset {\bf 1}^- : \ \ & & {\cal O} =
c \left( \begin{array}{ccc} 0 & 0 & 1 \\ 0 & 1 & 0 \\ 1 & 0 & 0
\end{array}
\right) .
\end{eqnarray}

\begin{table}
\begin{tabular}{c|ccccccc}
Sample element & $E$ & $R$ & $C_2, C_2 R$ & $C_3$ & $C_3^2$ & $C_3 R$
& $C_3^2 R$ \\ \hline
Order of class & 1 & 1 & 6 & 4 & 4 & 4 & 4 \\

Order of element & 1 & 2 & 4 & 6 & 3 & 3 & 6 \\ \hline  

${\bf 1}^0$ & 1 & 1 & 1 & 1 & 1 & 1 & 1 \\

${\bf 1}^+$ & 1 & 1 & 1 & $\eta$ & $\eta^2$ & $\eta$ & $\eta^2$ \\

${\bf 1}^-$ & 1 & 1 & 1 & $\eta^2$ & $\eta$ & $\eta^2$ & $\eta$ \\

${\bf 2}^0$ & 2 & $-2$ & 0 & 1 & $-1$ & $-1$ & 1 \\

${\bf 2}^+$ & 2 & $-2$ & 0 & $\eta$ & $-\eta^2$ & $-\eta$ & $\eta^2$ \\

${\bf 2}^-$ & 2 & $-2$ & 0 & $\eta^2$ & $-\eta$ & $-\eta^2$ & $\eta$ \\

${\bf 3}$ & 3 & 3 & $-1$ & 0 & 0 & 0 & 0
\end{tabular}
\caption{Character table of the double tetrahedral group $T'$.  The
phase $\eta$ is $\exp(2\pi i/3)$.}
\label{char}
\end{table}

\begin{table}
\begin{tabular}{c|cc}
SU(2) rep multiplicity & $T^\prime$ rep decomposition \\ \hline
$12N$ & $2N \left\{ {\bf 2}^0 \oplus {\bf 2}^+ \oplus
{\bf 2}^- \right\}$ & \\
$12N+1$ & ${\bf 1}^0 \oplus N \left\{ {\bf 1}^0 \oplus {\bf 1}^+
\oplus {\bf 1}^- \oplus 3 \cdot {\bf 3} \right\}$ & \\
$12N+2$ & ${\bf 2}^0 \oplus 2N \left\{ {\bf 2}^0 \oplus {\bf 2}^+ \oplus
    {\bf 2}^- \right\}$ & \\
$12N+3$ & ${\bf 3} \oplus N \left\{ {\bf 1}^0 \oplus {\bf 1}^+
\oplus {\bf 1}^- \oplus 3 \cdot {\bf 3} \right\}$ & \\
$12N+4$ & $\left\{ {\bf 2}^+ \oplus {\bf 2}^- \right\} \oplus 2N
    \left\{ {\bf 2}^0 \oplus {\bf 2}^+ \oplus {\bf 2}^- \right\}$ & \\
$12N+5$ & $\left\{ {\bf 1}^+ \oplus {\bf 1}^- \oplus {\bf 3} \right\}
\oplus N \left\{ {\bf 1}^0 \oplus {\bf 1}^+ \oplus {\bf 1}^- \oplus 3
\cdot {\bf 3} \right\}$ & \\
$12N+6$ & $(2N+1) \left\{ {\bf 2}^0 \oplus {\bf 2}^+ \oplus
{\bf 2}^- \right\}$ & \\
$12N+7$ & $\left\{ {\bf 1}^0 \oplus 2 \cdot {\bf 3} \right\} \oplus N
  \left\{ {\bf 1}^0 \oplus {\bf 1}^+ \oplus {\bf 1}^- \oplus 3 \cdot
  {\bf 3} \right\}$ & \\
$12N+8$ & ${\bf 2}^0 \oplus (2N+1) \left\{ {\bf 2}^0 \oplus {\bf 2}^+
\oplus {\bf 2}^- \right\}$ & \\
$12N+9$ & $\left\{ {\bf 1}^0 \oplus {\bf 1}^+ \oplus {\bf 1}^- \oplus
2 \cdot {\bf 3} \right\} \oplus N \left\{ {\bf 1}^0 \oplus {\bf 1}^+
\oplus {\bf 1}^- \oplus 3 \cdot {\bf 3} \right\}$ & \\
$12N+10$ & $\left\{ {\bf 2}^+ \oplus {\bf 2}^- \right\} \oplus (2N+1)
  \left\{ {\bf 2}^0 \oplus {\bf 2}^+ \oplus {\bf 2}^- \right\}$ & \\
$12N+11$ & $\left\{ {\bf 1}^+ \oplus {\bf 1}^- \oplus 3 \cdot {\bf 3}
\right\} \oplus N \left\{ {\bf 1}^0 \oplus {\bf 1}^+ \oplus {\bf 1}^-
\oplus 3 \cdot {\bf 3} \right\}$ &
\end{tabular}
\caption{Decomposition of SU(2) reps into reps of $T^\prime$.  $N$ is
  any nonnegative integer.}
\label{decomp}
\end{table}

\begin{table}
\begin{tabular}{l|l|l|l|l}
\multicolumn{5}{c}{$\epsilon = 0.04, \ \rho = 0.08, \ \epsilon^\prime
= 0.004, \ \xi = 0.017$} \\ \hline
$c_1 = -0.93 \pm 0.01$ & $d_1 = +1.33 \pm 0.45$ & $l_1 = +0.85 \pm
0.62$ & $r_1 = +0.94 \pm 0.84$ & $u_1 = +0.92 \pm 0.31$ \\
$c_2 = -0.46 \pm 0.03$ & $d_2 = -0.81 \pm 0.26$ & $l_2 = -1.01 \pm
1.11$ & $r_2 = +1.06 \pm 0.95$ & $u_2 = +1.48 \pm 0.70$ \\
$c_3 = -1.02 \pm 1.13$ & $d_3 = +1.55 \pm 0.67$ & $l_3 = -0.97 \pm
0.75$ & $r_3 = +1.03 \pm 1.12$ & $u_3 = -0.90 \pm 0.91$ \\
$c_4 = -1.03 \pm 1.15$ & $d_4 = +1.14 \pm 1.33$ & $l_4 = -1.09 \pm
1.04$ & $r_4 = -1.07 \pm 1.05$ & $u_4 =+1.07 \pm 1.21 $ \\
$c_5 = -0.90 \pm 0.01$ & $d_5 = -1.29 \pm 0.12$ & $l_5 = -1.11 \pm
0.79$ & $r_5 = -0.97 \pm 1.03$ & $u_5 = +1.84 \pm 0.95$ \\
$a = +0.98 \pm 1.06$ & & & &
\end{tabular}
\caption{Best fit parameters for the $T^\prime \times Z_3$ model with
$\tan \beta = 2$.  The minimum $\chi^2$ = 2.77.}
\label{fit}
\end{table}

\begin{table}
\begin{tabular}{lll}
Observable & Expt. value & Fit value \\ \hline
$m_u$ & $( 3.3 \pm 1.8 ) \times 10^{-3}$ & $3.5 \times 10^{-3}$ \\
$m_d$ & $( 6.0 \pm 3.0 ) \times 10^{-3}$ & $4.0 \times 10^{-3}$ \\
$m_s$ & $0.155 \pm 0.055$ & 0.136 \\
$m_c$ & $1.25 \pm 0.15$ & 1.24 \\
$m_b$ & $4.25 \pm 0.15$ & 4.25 \\
$m_t$ & $173.8 \pm 5.2$ & 170.4 \\
$m_e$ & $( 5.11 \pm 1\% ) \times 10^{-4}$ & $5.11 \times 10^{-4}$ \\
$m_\mu$ & $0.106 \pm 1\%$ & 0.106 \\
$m_\tau$ & $1.78 \pm 1\%$ & 1.78 \\
$|V_{us}|$ & $0.221 \pm 0.004$ & 0.221 \\
$|V_{ub}|$ & $( 3.1 \pm 1.4 ) \times 10^{-3}$ & $2.3 \times 10^{-3}$
\\
$|V_{cb}|$ & $( 3.9 \pm 0.3 ) \times 10^{-2}$ & $3.9 \times 10^{-2}$
\\
$\Delta m_{23}^2 / \Delta m_{12}^2$ & 100 -- 2500 & 526 \\
$\ln \left( \Delta m_{23}^2 / \Delta m_{12}^2 \right)$ & $6.22 \pm
1.61$ & 6.27 \\
$\sin^2 2\theta_{12}$ & $2 \times 10^{-3}$ -- $10^{-2}$ & $4.5 \times
10^{-3}$ \\
$\ln \left( \sin^2 2\theta_{12} \right)$ & $-5.41 \pm 0.80$ & $-5.40$
\\
$\sin^2 2\theta_{23}$ & $> 0.8$  & 0.90 \\
$\sin^2 2\theta_{13}$ & --- & $1.4 \times 10^{-3}$
\end{tabular}
\caption{Experimental values versus fit central values for observables
using the inputs of Table~\ref{fit}.  Masses are in GeV and all other
quantities are dimensionless. Error bars indicate the larger of 
experimental or 1\% theoretical uncertainties, as described in
the text.}
\label{data}
\end{table}

\begin{figure}
  \begin{centering}
  \def\epsfsize#1#2{1.0#2}
\hfil\hspace{-10em} \epsfbox{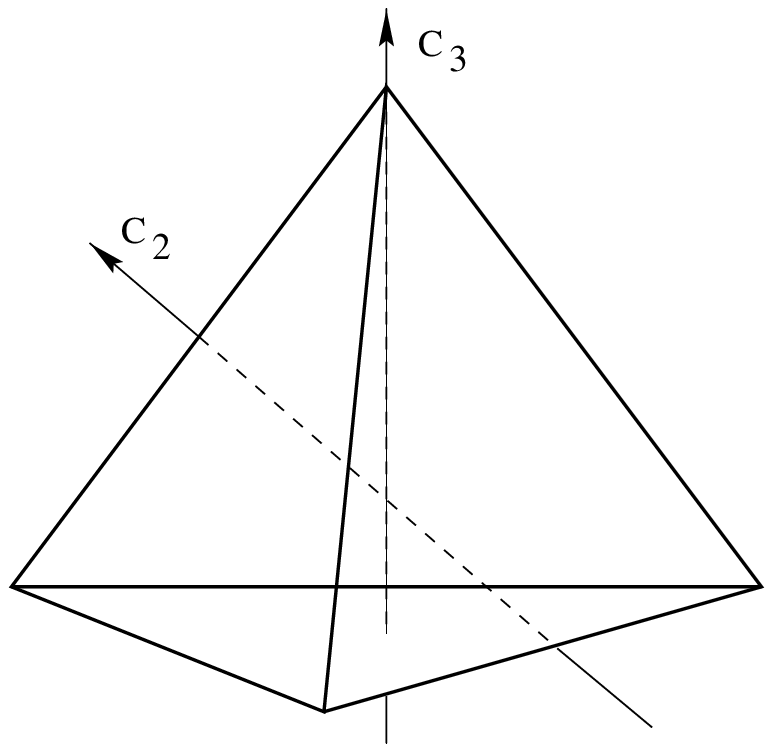} \hfill
\caption{Geometrical illustration of the group $T'$ or $T$.  The rotations
$C_2$ and $C_3$ generate all other rotations in each group.}
  \end{centering}
\end{figure}


\begin{thebibliography}{99}

\bibitem{DN}M. Dine, A.E. Nelson, Phys.\ Rev.\ D {\bf 48}, 1277 (1993);
M. Dine, A.E. Nelson, Y. Shirman, Phys.\ Rev.\ D {\bf 51}, 1362 (1995).

\bibitem{randall}L. Randall and R. Sundrum, Nucl.\ Phys.\ {\bf B557}, 79 (1999).

\bibitem{luty}G.F. Giudice, M.A. Luty, H. Murayama, and R. Rattazzi,
JHEP {\bf 9812}, 027 (1998)

\bibitem{dim}I. Antoniadis, S. Dimopoulos, A. Pomarol, M. Quiros, 
 Nucl.\ Phys.\ {\bf B544}, 503 (1999).

\bibitem{abelian} See, for example: \\
Y. Nir and N. Seiberg, Phys.\ Lett.\ B {\bf 309}, 337 (1993);
M. Leurer, Y. Nir, and N. Seiberg, Nucl.\ Phys.\ {\bf B420}, 468;
L.E. Ib\'{a}\~{n}ez and G.G. Ross, Phys.\ Lett.\ B {\bf 332}, 100 (1994);
P. Binetruy and P. Ramond, {\it ibid}. {\bf 350}, 49 (1995);
V. Jain and R. Shrock, {\it ibid.} {\bf 352}, 83 (1995); Stony Brook
Report No.\ ITP-SB-95-22, hep-ph/9507238 (unpublished);
Y. Nir, Phys.\ Lett.\ B {\bf 354}, 107 (1995);
E. Dudas, S. Pokorski, and C.A. Savoy, {\it ibid.} {\bf 356}, 45
(1995);
M.M. Robinson and J. Ziabicki, Phys.\ Rev.\ D {\bf 53}, 5924
(1996);
E. Dudas, S. Pokorski, and C.A. Savoy, Phys.\ Lett.\ B {\bf 369}, 255
(1996);
A. Pomarol and D. Tommasini, Nucl.\ Phys.\ {\bf B466}, 3 (1996).

\bibitem{nonabelian} See, for example: \\
D.B. Kaplan and M. Schmaltz, Phys.\ Rev.\ D {\bf 49}, 3741 (1994); 
M. Dine, R. Leigh, and A. Kagan, Phys.\ Rev.\ D {\bf
48}, 4269 (1993); L.J. Hall and H. Murayama, Phys.\ Rev.\ Lett.\ {\bf
75}, 3985 (1995); C.D. Carone, L.J. Hall, and H. Murayama, 
Phys.\ Rev.\ D {\bf 53}, 6282 (1996);  P.H. Frampton and O.C.W. Kong, 
Phys.\ Rev.\ Lett.\ {\bf 77}, 1699 (1996); P.H. Frampton 
and A. Ra\v{s}in, hep-ph/9910522
(unpublished).

\bibitem{arason} See, for example,
H. Arason, D.J. Casta\~{n}o, E.J. Piard, and P. Ramond, 
Phys.\ Rev.\ D {\bf 47}, 232, (1993).
\bibitem{u21}
R. Barbieri, G. Dvali, and L.J. Hall,  Phys.\ Lett.\ B {\bf 377}, 76 (1996).

\bibitem{u22}
R. Barbieri, L.J. Hall, and A. Romanino, Phys.\ Lett.\ B {\bf 401}, 47
(1997).

\bibitem{u23}
R. Barbieri, L.J. Hall, S. Raby and A. Romanino, Nucl.\ Phys.\ {\bf B493}, 
3 (1997).

\bibitem{GJ}
H. Georgi and C. Jarlskog, Phys.\ Lett.\ B {\bf 86}, 297 (1979).

\bibitem{acl} 
A. Aranda, C.D. Carone, R.F. Lebed, hep-ph/9910392 (To appear in Phys.\ Lett.\ B).

\bibitem{worm}S. Coleman, Nucl.\ Phys.\ {\bf B310}, 643 (1988);\\
S. Giddings and A. Strominger, {\it ibid.} {\bf B307}, 854 (1988);\\
G. Gilbert, {\it ibid.} {\bf B328}, 159 (1989).

\bibitem{GS}
M. Green and J. Schwarz, Phys.\ Lett.\ B {\bf 149}, 117
(1984).

\bibitem{barcre}
R. Barbieri, P. Creminelli, A. Romanino, Nucl.\ Phys.\ {\bf B559}, 
17 (1999).

\bibitem{raby} 
R. Dermisek, S. Raby, hep-ph/9911275 (unpublished).
  
\bibitem{banks}T. Banks, Nucl.\ Phys.\ {\bf B323}, 90 (1989).

\bibitem{KW}L.M. Krauss and F. Wilczek, Phys.\ Rev.\ Lett.\ {\bf 62},
1221 (1989).

\bibitem{houches}J. Preskill, {\it P. Ramond and R. Stora, eds Les
Houches, Session XLIV, 1985-Architecture des interactions
fondamentales} $\grave{a}$ {\it courte distance/Architecture of fundamental
interactions at short distances}, 235, Elsevier Science Publishers B.V., 1987

\bibitem{IR} L.E. Ib\'{a}\~{n}ez and G.G. Ross, Phys.\ Lett.\ B {\bf
260}, 291 (1991).

\bibitem{BD} T. Banks and M. Dine, Phys.\ Rev.\ D {\bf 45}, 1424
(1992);
J. Preskill, S.P. Trivedi, F. Wilczek, and M.B. Wise, Nucl.\ Phys.\
{\bf B363}, 207 (1991).

\bibitem{d6}
C.D. Carone and R.F. Lebed, Phys.\ Rev.\ D {\bf 60}, 096002 (1999). 

\bibitem{TW} A.D. Thomas and G.V. Wood, {\it Group Tables}, Shiva
Publishing, Orpington, UK, 1980.

\bibitem{barhall}
R. Barbieri, L.J. Hall, D. Smith, A. Strumia and N. Weiner, JHEP 
{\bf 9812}, 017, (1998).

\bibitem{Kam}Super-Kamiokande Collaboration (Y. Fukuda et al.),
Phys.\ Rev.\ lett. {\bf 81}, (1998) 1562.

\bibitem{bah}J.N. Bahcall, P.I. Krastev, A.Y. Smirnov,
Phys.\ Rev.\ D {\bf 58}, 096016 (1998).

\bibitem{BBO}V. Barger, M.S. Berger, P. Ohmann, Phys.\ Rev.\ D {\bf 47},
1093 (1993).

\bibitem{PDG}Particle Data Group (C. Caso et al.), 
Eur.\ Phys.\ J.\ C {\bf 3}, 1 (1998).

\bibitem{AKL}I. Antoniadis, C. Kounnas, R. Lacaze, Nucl.\ Phys.\
{\bf B211}, 216 (1983).

\bibitem{NRF}W.H. Press, S.A. Teukolsky, W.T. Vetterling, B.P. Flannery,
{\it Numerical Recipes in Fortran 77}, 2nd ed. (Cambridge University Press,
New York, 1992).

\bibitem{BLP}K.S. Babu, C.N. Leung, J. Pantaleone, Phys.\ Lett.\ B {\bf
319}, 191 (1993). 

\bibitem{mas}
F. Gabbiani, E. Gabrielli, A. Masiero and L. Silvestrini,
Nucl.\ Phys.\ {\bf B477}, 321 (1996).

\bibitem{nelson}A.G. Cohen, D.B. Kaplan, A.E. Nelson,
Phys.\ Lett.\ B {\bf388}, 588 (1996). 
\bibitem{caronehall}
C.D. Carone and L.J. Hall, Phys.\ Rev.\ D {\bf 56}, 4198 (1997).

\bibitem{hallweiner}
L.J. Hall and N. Weiner, Phys.\ Rev.\ D {\bf 60}, 033005 (1999).  

\bibitem{lsnd}
LSND Collaboration, Phys.\ Rev.\ Lett.\ {\bf 81}, 1774 (1998).

\end{thebibliography}
\end{document}